\documentclass{aa}  

\usepackage{graphicx}
\usepackage{txfonts}
\usepackage[version=3]{mhchem}
\usepackage{xcolor}
\usepackage{lscape}

\usepackage{booktabs}
\usepackage{hyperref}
\usepackage{ulem}
\usepackage{csquotes}
\newcommand{\micron}{\textmu m\xspace}

\begin{document}

   \title{Fate and detectability of rare gas hydride ions in nova ejecta}

   \subtitle{A case study with nova templates}

   \author{M. Sil\inst{\ref{inst1},\ref{inst2},\ref{inst3}} \and A. Das\inst{\ref{inst4},\ref{inst7}} \and R. Das\inst{\ref{inst3}} \and R. Pandey\inst{\ref{inst5}} \and A. Faure\inst{\ref{inst1}} \and H. Wiesemeyer\inst{\ref{inst6}} \and P. Hily-Blant\inst{\ref{inst1}} \and F. Lique\inst{\ref{inst2}} \and P. Caselli\inst{\ref{inst7}}}

   \institute{Univ. Grenoble Alpes, CNRS, IPAG, 38000 Grenoble, France;
              \email{milansil93@gmail.com}
              \label{inst1}
        \and
             Univ Rennes, CNRS, IPR (Institut de Physique de Rennes) - UMR 6251, F-35000 Rennes, France
             \label{inst2}
        \and
             S. N. Bose National Centre for Basic Sciences, Block-JD, Sector-III, Salt Lake, Kolkata 700106, India
             \label{inst3}
         \and
             Institute of Astronomy Space and Earth Science, P177 CIT Road, Scheme 7m, Kolkata 700054, India
             \label{inst4}
         \and
             Astronomy \& Astrophysics Division, Physical Research Laboratory, Ahmedabad 380009, Gujarat, India
             \label{inst5}
         \and
             Max-Planck-Institut für Radioastronomie, Auf dem Hügel 69, D-53121 Bonn, Germany
             \label{inst6}
         \and
             Max-Planck-Institute for extraterrestrial Physics, P.O. Box 1312 85741 Garching, Germany
             \label{inst7}}


 
  \abstract 
{HeH$^+$ was the first heteronuclear molecule to form in the metal-free Universe after the Big Bang. The molecule gained significant attention following its first circumstellar detection in the young and dense planetary nebula NGC 7027.
We target some hydride ions associated with the noble gases (HeH$^+$, ArH$^+$, and NeH$^+$) to investigate their formation in harsh environments like the nova outburst region.
We use a photoionization modeling (based on previously published best-fit physical parameters) of the moderately fast ONe type nova, QU Vulpeculae 1984, and the CO type novae, RS Ophiuchi and V1716 Scorpii.
Our steady-state modeling reveals a convincing amount of HeH$^+$, especially in the dense clump of RS Ophiuchi and V1716 Scorpii. The calculated upper limit on the surface brightness of HeH$^+$ transitions suggests that the \textit{James Webb} Space Telescope (JWST) could detect some of them, particularly in sources like RS Ophiuchi and V1716 Scorpii, which have similar physical and chemical conditions and evolution.
It must be clearly noted that the sources studied are used as templates, and not as targets for observations.
The detection of these lines could be useful for determining the physical conditions in {similar types of systems and for validating our predictions based on new electron-impact ro-vibrational collisional data at temperatures of up to 20\,000~K}.}


   \maketitle

\section{Introduction} \label{sec:intro}
Classical novae (CNe) are catastrophic events that occur on the surface of a white dwarf (WD) that accretes hydrogen-rich matter from a companion star, typically a main sequence star or a red giant star.
The WD is generally a carbon-oxygen (CO) or oxygen-neon (ONe) type, whereas the secondary generally is a K or M main sequence star \citep{jose12}. The outburst occurs due to thermonuclear runaway (TNR) on the WD surface when the accreted material reaches a critical temperature and pressure.
The explosion releases a significant amount of energy ($\sim10^{40}-10^{45}$~erg~s$^{-1}$) and ejects a shell with a material mass of $\sim10^{-6}-10^{-4}$~M$_\sun$ \citep{gehr98} with velocities ranging from several hundred to $\ge1000$~km~s$^{-1}$ \citep[][and references therein]{bode08}.
Sometime after the explosion, the accretion process begins, and the system prepares for the next outburst.
The outburst ejects matter into the surrounding area without disrupting the WD while continuing accretion for subsequent outburst periods ranging from tens to hundreds of years. Under specific conditions, such as those related to certain properties of the WD and the accretion process, these outbursts occur on timescales ranging from years to decades, indicating an accelerated rate of mass transfer within the binary system. This categorizes them as recurrent novae (RNe). Symbiotic novae (SymNe) represent a subclass of RNe characterized by systems with significant orbital separations and a companion red giant star \citep{starrfield_2020}.

Following an outburst, a small fraction of observed novae have been found to form dust and molecules inside the ejecta. The first observed molecule is CN, which was detected in the optical spectrum of nova DQ Her (1934) by \cite{wils35} shortly after the eruption.
At much later epochs, the 2.122~\micron transition of \ce{H_2} was detected in the DQ Her remnant \citep{evan91}.
\cite{ferl79} identified the most commonly detected emission features of CO at 4.8~\micron (fundamental bands) and 2.3~\micron (first overtone bands) in NQ Vul 1976.
\cite{rawl88} deduced a CO column density of $10^{17}-10^{18}$~cm$^{-2}$ (corresponding to 1~--~10\% of all available carbon in the ejecta being locked up in CO), indicating the high CO formation efficiency for NQ Vul. 
Subsequently, the first-overtone CO emission was observed in many novae; for example V842 Cen \citep[Nova Cen 1986;][]{hyla89,wich90}, Cas 1993 \citep[V705 Cas;][]{evan96}, V2274 Cyg \citep{rudy03}, and V2615 Oph \citep{das09}. 
Apart from CO, in a mid-infrared spectroscopic survey of novae, \cite{smit95} tentatively identified the $\upsilon=2-1$ transition of SiO in V992 Sco (1992) at an excitation temperature of 1500~K.
An Al-bearing molecule, AlO, was tentatively identified in the spectrum of GK Per \citep[1901;][]{bian86} and RS Car \citep[1895;][]{bian01}.
The transient absorption bands of both C$_2$ and CN were detected in the optical wavelength around the visual brightness maximum of the CNe V2676 Oph \citep{2015PASJKawakita} and V1391 Cas \citep{2021ApJFuji}.

Despite significant research over decades, dust formation in nova ejecta has remained an enigma \citep[][and references therein]{rawl88,1989MNRASRawlings,1994NaturShore,2004MNRASPontefract,2004A&AShore,shor18}. Recent studies have proposed the idea of \enquote{internal shocks} occurring within the nova ejecta. These shocks provide favorable circumstances for the creation of dust in the ejecta. The cooling efficiency of the gas in the shock zones causes an increase in the ejecta density, creating clumps of cold, neutral, and dense gas. These dense regions are well protected from the intense radiation emitted by the central ionizing WD \citep[see, e.g.,][and references therein]{2014NaturChomiuk,2014MNRASMetzger}. \cite{Derszinski_2017} proposed a dust-shock model in which the powerful radiative shocks within the nova ejecta create very dense and cool layers of gas behind these shocks, providing favorable conditions for the formation of dust nucleation sites. Nova V2891 Cyg provided most likely the first observational evidence of such shock-induced dust formation within nova ejecta \citep{Kumar_2022,Pandey2024}. In a recent study, \cite{bane23} detected CO and dust in the recurrent nova V745 Sco about 8.7 days after its 2014 outburst. In the ejecta of a recurrent nova, this is the first detection of molecules or dust, which have probably formed in a cool, dense, and clumpy region between the forward and reverse shocks.

In the present work, we investigate the possibility of the formation of the noble gas hydrides (ArH$^+$, NeH$^+$, and HeH$^+$) in the outburst of two different types of novae, namely CO type and ONe type, and discuss the future detection of these hydrides in similar types of nova ejecta with a space-based telescope such as the \textit{James Webb} Space Telescope (JWST).
Hydrides are the first molecules to form in any astrophysical environment, as hydrogen is the most abundant species.
Noble gas hydrides have already been observed in different radiation-dominated regions such as the Crab nebula \citep[ArH$^+$;][]{barl13} and the NGC 7027 planetary nebula \citep[HeH$^+$;][]{gust19,neuf20}. Recently, \cite{das20} studied the chemistry of noble gas hydride and hydroxyl cations in the Crab filamentary region and found a favorable parameter space (hydrogen number density of $\sim 10^{4-6}$~cm$^{-3}$ 
and a cosmic-ray ionization rate per H$_2$ of 
$\sim 10^{-11} - 10^{-10}$~s$^{-1}$) to explain the observed features.

For the present study, we selected three different novae from the literature that have been investigated using photoionization modeling: QU Vulpeculae 1984 (hereafter, QU Vul), RS Ophiuchi (hereafter, RS Oph), and V1716 Scorpii (hereafter, V1716 Sco). It should be noted that these sources did not exhibit dust formation following an outburst. Other well-studied dust-forming CNe, such as V705 Cas, V723 Cas, and V5668 Sgr, might be more appropriate for studying molecule formation due to their unique characteristic of forming dust post-outburst \citep{gehr18,shor18}. However, modeling the formation of noble gas hydrides in these sources is beyond the scope of this study because photoionization model studies are not available for them in the literature.
In our study, we focus on those novae whose best-fit photoionization model parameters are present in the literature, as we adopt these published results as preliminary parameters for our modeling.

QU Vul is a moderately fast nova with a ONeMg WD \citep{livi94} that has a mass of $\sim$0.82~--~0.96~M$_\sun$ \citep{hach16}. QU Vul was discovered by \cite{coll84} and further confirmed spectroscopically by \cite{rosi87}.
Using multiwavelength spectroscopic data at various epochs during the nebular phase of outbursts, \cite{schw02} 
estimated the best-fit elemental abundances in the ejecta of QU Vul.

RS Oph is a well-studied Galactic symbiotic RNe with an average recurrence timescale of about 15 years \citep{scha10,pand22}.
RS Oph, which is a probable Type Ia supernova progenitor, has a massive CO-type WD with a mass range of 1.2~--~1.4~M$_\sun$ \citep{miko17} and an evolved red giant companion with a mass of 0.68~--~0.80~M$_\sun$ \citep{bran09}. 
The RS Oph system is renowned for generating shock waves \citep{hjel86,soko06}. These shock waves occur when the material ejected from the WD of RS Oph interacts with the surrounding winds of its red giant companion. As a result, a shock wave travels through the red giant wind, causing the shock velocity to decrease gradually over time.

V1716 Sco, also known as PNV J17224490-4137160, was first discovered in outburst by A. Pearce\footnote{\url{http://www.cbat.eps.harvard.edu/unconf/followups/J17224490-4137160.html}} on 2023 April 20.6780 UT at an unfiltered magnitude of 8.0. \citet{2023ATel16003Walter} classified it as a Fe\,{\sc ii}-type CNe by examining the spectral evolution. The time it takes for the brightness of the object to decline by 2 magnitudes (t$_2$) and 3 magnitudes (t$_3$) from its maximum was estimated to be around 5.8 and 11.7 days, respectively, making it a fast nova \citep{wood24}. The distance of the nova is estimated to be around $D = 3.6 \pm 0.6$ kpc \citep{wood24}. V1716 Sco was also observed in gamma-rays above 100~MeV by Fermi. The X-ray observations using \textit{swift} revealed that the nova transitioned into a supersoft X-ray (SSS) phase around day 55 \citep{2023ATelPage}.

We performed photoionization modeling {(based on the earlier published best-fit physical parameters)} to estimate the intensity of the noble gas hydrides in the spectra of the novae.
We use new electron-impact ro-vibrational collisional data for HeH$^+$ at temperatures of up to 20\,000 K (see \autoref{sec:appendix_1}), the reaction network prepared by \cite{das20} (see \autoref{table:reaction_Ar}) for this study, and the best-fit model parameters (see Table \ref{tab:phy}) estimated by \cite{schw02} for QU Vul, \cite{pand22} for RS Oph, and \cite{wood24} for V1716 Sco.
The paper is organized as follows.  Section~\ref{sec:physical_cond} describes the physical and chemical properties implemented in the photoionization model. Section~\ref{sec:results} discusses the obtained modeling results for the three different novae. Finally, we provide concluding remarks in Sect.~\ref{sec:conclusions}.


\section{Physico-chemical conditions} \label{sec:physical_cond}

To model the nova ejecta, we used the latest version of the photoionization code \textsc{Cloudy}, v23.01 \citep[][and references therein]{chat23}.
\textsc{Cloudy} is a spectral synthesis code that simulates physical conditions over a broad range of physical parameters (temperature, density, radiation field, etc.).
Thermal and statistical equilibrium equations, charge, and energy conservation are simultaneously solved in \textsc{Cloudy} \citep{oste06}. The ionization level is calculated by balancing the most important ionization (photo, Auger, collisional ionization, and charge transfer) and recombination processes (low-temperature dielectronic, high-temperature dielectronic, three-body recombination, radiative, {dissociative}, and charge transfer). The free electrons are considered to belong to a predominantly Maxwellian velocity distribution. The kinetic temperature is determined by the balance between heating (mechanical, photoelectric, cosmic-ray, etc.) and cooling (mainly inelastic collisions between the electrons and other particles) processes. The associated continuum and line radiative transfer are solved simultaneously. The predicted flux of emission lines can be directly compared to the measured line fluxes of the observed spectra.

\begin{table*}
\caption{\textsc{Cloudy} physical model parameters and initial elemental abundance set for QU Vul epoch D550 \citep{schw02}, RS Oph epoch D22 \citep{pand22}, and {V1716 Sco epoch D132.8 \citep{wood24}}.
\label{tab:phy}}
\centering
\begin{tabular}{cccc}
    \hline
{\bf Parameter} & {\bf QU Vul D550} & {\bf RS Oph D22} & {\bf V1716 Sco D132.8} \\
    \hline
Temperature of central ionizing WD $\rm{T_{BB}}$ (K) & $3.2\times10^5$ & $6.60\times10^4$ & {$5.5\times10^5$} \\
Luminosity of central ionizing WD (erg s$^{-1}$) & $3.7\times10^{37}$ &  $1.0\times10^{37}$ & {$3.16\times10^{37}$} \\
Hydrogen density, $n(H)$ (cm$^{-3}$) & $5.6\times10^6$ & $0.35\times10^{10a},\ 1.77\times10^{8b}$ & {$7.08\times10^{7a},\ 2.51\times10^{6b}$} \\
Inner radius, $r_{in}$ (cm) & $4.7\times10^{15}$ & $3.16\times10^{14}$ & {$1.02\times10^{15}$} \\
Outer radius, $r_{out}$ (cm) & $2.3\times10^{16}$ & $5.49\times10^{14}$ & {$2.17\times10^{15}$} \\
{Cylinder height, $h$ (cm)} & --- & --- & {$2.24\times10^{15}$} \\
$\alpha$ & $-3$ & $-3$ & {$-3$} \\
Covering factor & 1.0 & $0.61^a,\ 0.39^b$ & {$0.5$} \\
Filling factor, $f(r)$ & 0.10 & 0.01 & {$0.20^a,\ 0.80^b$} \\
$\beta$ & 0.0 & 0.0 & {0.0} \\
Galactic cosmic-ray ionization rate, {$\zeta_{H_2}$} (s$^{-1}$) & $4.6 \times 10^{-16}$ & $4.6 \times 10^{-16}$ & {$4.6 \times 10^{-16}$} \\
\hline
\multicolumn{4}{c}{\bf Initial elemental abundance (with respect to total hydrogen nuclei) set} \\
\hline
{\bf Species} & {\bf QU Vul D550$^*$} & {\bf RS Oph D22$^*$} & {\bf V1716 Sco D132.8} \\
    \hline
{He} & 1.16 & 1.9 & {$1.87\times10^{-1}$} \\
{C} & 0.2 & --- & {$1.74\times10^{-3}$} \\
{N} & 12 & 95 & {$1.67\times10^{-2}$} \\
{O} & 2.5 & 2.6 & {$9.31\times10^{-3}$} \\
{Ne} & 21.7 & --- & {$2.18\times10^{-4}$} \\
{Mg} & 10 & --- & {$1.19\times10^{-4}$} \\
{Al} & 53.3 & --- & {$1.69\times10^{-5}$} \\
{Si} & 2.0 & --- & {$7.13\times10^{-5}$} \\
{Ar} & 0.28 & --- & {---}\\
{Fe} & 0.53 & 1.0 & {$6.32\times10^{-5}$} \\
{S} & --- & --- & {$9.90\times10^{-5}$} \\
{Ca} & --- & --- & {$5.26\times10^{-6}$} \\
{P} & --- & --- & {$3.08\times10^{-5}$} \\
\hline
\end{tabular}
\tablefoot{
The initial elemental abundance set for QU Vul is considered by taking an average over all four epochs \citep[D310, D470, D550, and D630;][]{schw02}. \\
{The initial elemental abundances of the ejecta (only those that were observed) are noted. All other elements (without observed lines) are kept at their solar abundances \citep{grev93,grev10}.} \\
{$^*$ Relative to solar values.} \\
$^a$ High-density ("clump") component of ejecta. \\
$^b$ Low-density ("diffuse") component of ejecta.
}
\end{table*}

We consider a central ionizing source to have a blackbody ($\rm{T_{BB} > 10^4}$~K) surrounded by a spherically symmetric expanding shell located at a distance of $r_{in}$ (inner radius) away from the central source. The thickness of the shell is set by specifying an outer radius ($r_{out}$).
Following \cite{bath76}, the density in the ejecta is of the form
\begin{equation}
    n(r) = n(H)(r/r_{in})^\alpha ,
\end{equation}
where $n(r)$ and $n(H)=n(r_{in})$ represent the total hydrogen nuclei density (in all forms, e.g., neutral, ionized, molecular, and
any other) of the ejecta at a distance $r$ and inner radius ($r_{in}$), respectively, and $\alpha$ is the power-law index.
Assuming the surrounding gas to be clumpy, the clumpiness is considered by setting the radial-dependent filling factor, $f(r)$ as,
\begin{equation}
    f(r) = f(r_{in})(r/r_{in})^\beta ,
\end{equation}
where $\beta$ is the exponent of the power law. The filling factor describes the fraction of the volume occupied by the gas (i.e., clumpiness).
The initial elemental abundances of the ejecta (only those that were observed) are noted in \autoref{tab:phy}. At the same time, all other elements (without observed lines) are kept at their solar abundances \citep{grev93,grev10}.

The best-fit physical parameters are taken from \cite{schw02}, \cite{pand22}, and \cite{wood24} for QU Vul, RS Oph, and V1716 Sco, respectively (see \autoref{tab:phy}).
The chemical reaction network for HeH$^+$, ArH$^+$, and NeH$^+$ (excluding the reaction pathways of the hydroxyl
cations, i.e., ArOH$^+$, NeOH$^+$, and HeOH$^+$) is adopted from \cite{das20}, who considered the rate coefficients based on the available literature \citep{schi14,prie17,gust19,neuf20} and databases \citep{mcel13}.
Additionally, the recently measured temperature-dependent dissociative recombination (DR) rate coefficient of ArH$^+$ \citep{kalo24} is included in our network.
We note that \cite{kalo24} provided a set of piecewise-joined fit functions on several temperature intervals because a single Arrhenius-Kooij fit function could not adequately model their experimental results. This approach introduces discontinuities in the temperature dependence of the analytical rate coefficient between the temperature intervals. We have chosen a specific temperature interval of $7000-20\,000$~K from their Table~III that is suitable for the conditions in the nova ejecta regions. To run the \textsc{Cloudy} code for the chemistry at temperatures ranging from 10 to 20\,000~K, we ensured that the poor fit below 7000~K does not affect the analysis of the nova ejecta in the three sources we are studying. We assume the same temperature-dependent DR rate coefficient for NeH$^+$ as that of ArH$^+$.
The detailed list of reactions and the corresponding rate coefficients adopted here {(for HeH$^+$, ArH$^+$, and NeH$^+$)} are presented in \autoref{sec:appendix_3} and discussed in detail in \cite{das20}.
All other species, including neutral atoms, molecules, and ions, and their corresponding reactions, are kept similar to the default settings in \textsc{Cloudy}, v23.01.
{It is important to note that electronic recombination (ER) reactions of all the atomic ions, ionization by direct X-rays ($\zeta_{XR}$), ionization by secondary photons from X-rays ($\zeta_{XRPHOT}$), and electron-impact X-ray ionization ($\zeta_{XRSEC}$) for all basic elements and atomic ions are self-consistently controlled in \textsc{Cloudy} by default. We list them in \autoref{table:reaction_Ar} for the sake of completeness.}

{The Galactic cosmic-ray background with a mean primary cosmic-ray ionization rate per hydrogen atom ($\zeta_H$) of $2\times10^{-16}$~s$^{-1}$ \citep{indr07} is considered for all the models. Cosmic-ray ionization rates of molecular hydrogen ($\zeta_{H_2}$) of $4.6\times10^{-16}$~s$^{-1}$ are self-consistently determined in \textsc{Cloudy} following the approach and conversion factor given by \cite{glas74}.}
The present background radiation temperature is assumed to be $2.725 \pm 0.002$~K. Following \cite{shaw05}, an extensive model of the H$_2$ molecule is considered.
Considering a closed spherical shell geometry, the chemistry and thermal balance at each depth are calculated self-consistently.

We consider the collisional data files for HeH$^+$, {$\rm{^{36}ArH^+}$}, and NeH$^+$ for calculating the surface brightness using \textsc{Cloudy}, which are discussed in Sect.~\ref{sec:results}.
The collisional data file for HeH$^+$ is unavailable in the \textsc{Cloudy} database, and therefore the collisional data file for HeH$^+$ is considered from the Excitation of Molecules and Atoms for Astrophysics (EMAA) database\footnote{\url{https://emaa.osug.fr} and \url{https://dx.doi.org/10.17178/EMAA}} \citep{hami16,desr20}\footnote{\url{https://dx.doi.org/10.17178/EMAA_HeH-plus_Rotation_2ccc89ee}}, which includes 12 rotational energy levels in the range $J=0-11$ {at ground vibrational level $\upsilon=0$}.
The collisional data file for HeH$^+$ contains the collisional rates with two collision partners: electrons (12 temperatures in the range 10~--~3000~K) and atomic hydrogen (18 temperatures in the range 10~--~500~K).
The temperature of the nova outburst region is much higher than the temperature range provided in the HeH$^+$ data file.
To have realistic collisional rates for the nova outburst region,
the Coulomb-Born (CB) approximation (see \autoref{sec:appendix_1} for detailed discussion) is used to compute the electron-impact ro-vibrational rate coefficients up to $\sim$20\,000~K, which are suitable for the nova outburst region.
We extract the radiative rates from the \texttt{EXOMOL} database \citep{exomol16,amar19} to prepare the spectroscopy part. The new data file includes 162 bound ro-vibrational energy levels, with the highest level being $\upsilon=11, \ J=1,$ which lies 14\,873.7~cm$^{-1}$ (about 21\,400~K) above the ground level. Quasi-bound levels have large dissociation probabilities and were neglected. Only electrons are used as the colliding partner in this extended data file as the electron abundance is significantly larger than other collisional partners (e.g., H) ---this is discussed below. We calculate the rates at 21 temperatures within the temperature range of 10~--~20\,000~K. {We consider 1282 radiative and 553 collisional} ($\Delta \upsilon=0, \Delta J=\pm 1, \pm 2$ and $\Delta \upsilon=\pm 1, \Delta J=\pm 1$) transitions in this file.  This newly generated data file is used as default throughout this work, unless stated otherwise.

We replace the collisional data file for {$\rm{^{36}ArH^+}$} (already available in the \textsc{Cloudy} \texttt{data/lamda} directory considering the hydrogen atom as the only collision partner) with that available in the EMAA database\footnote{\url{https://dx.doi.org/10.17178/EMAA_(36Ar)H-plus_Rotation_173ef621}}. This latter includes 12 rotation energy levels in the range $J=0-11$ {at ground vibrational level $\upsilon=0$} and contains the collisional rates with two collision partners: electrons \citep[for 12 temperatures in between 10~--~3000~K;][]{hami16} and atomic hydrogen \citep[for 34 temperatures within the range of 10~--~1000~K;][]{dagd18}.
For NeH$^+$, no collisional rates are available. 
Thus, we approximate the collisional rates {of NeH$^+$} by considering the collisional rates of {$\rm{^{36}ArH^+}$} with electrons as the only colliding partners \citep[for nine temperatures within the range of 5~--~3000~K;][]{hami16}.
The collisional data file for NeH$^+$ includes eight rotation energy levels in the range $J=0-7$ at ground vibrational level $\upsilon=0$. There are 7 radiative and 28 collisional transitions.
We note that due to the unavailability of high-temperature data for {$\rm{^{36}ArH^+}$} and NeH$^+$, some predictions may be inaccurate (see \autoref{sec:appendix_2}).


\section{Results and discussion} \label{sec:results}

\subsection{QU Vul} \label{sec:qu_vul}

The epoch D550 for QU Vul is chosen here since it contains the greatest number of observed emission lines \citep{schw02}. The D550 epoch considered the average optical line ratios for the days 530 and 560 along with the day 554 UV lines.
The initial elemental abundances and physical parameters used for the model are listed in \autoref{tab:phy}. These are the best-fit parameters obtained by \cite{schw02}.
The obtained geometry is both spherical and closed.

\begin{figure}
    \centering
    \includegraphics[width=0.49\textwidth]{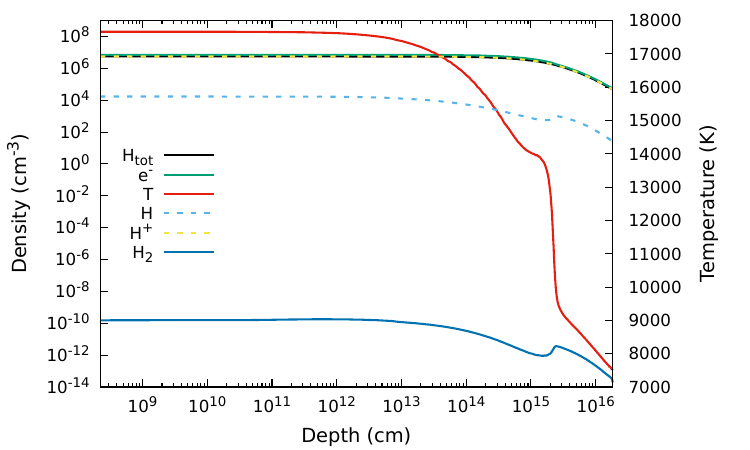}
    \caption{Density and temperature profile in QU Vul ejecta.}
    \label{fig:phy_qu_vul}
\end{figure}

\autoref{fig:phy_qu_vul} depicts the density and temperature variations throughout the depth ($dr = r_{out}
- r_{in}$), that is, the distance starting from the illuminated face of the cloud ($r_{in}$) through the entire thickness of the  shell to the outer edge ($r_{out}$).
The electron temperature varies in the range of 7520~--~17\,700~K.
{The very high density of electrons and H$^+$ signifies that the shell is in an ionized form, and that the electrons are much more abundant than atomic hydrogen. We obtain a peak electron density of $7.1\times10^6$~cm$^{-3}$, whereas it is $1.6\times10^4$~cm$^{-3}$ for atomic hydrogen.} Therefore, the collisional rate with electrons should dominate for the non-local thermodynamic equilibrium emission.

\begin{figure}
    \centering
    \includegraphics[width=0.49\textwidth]{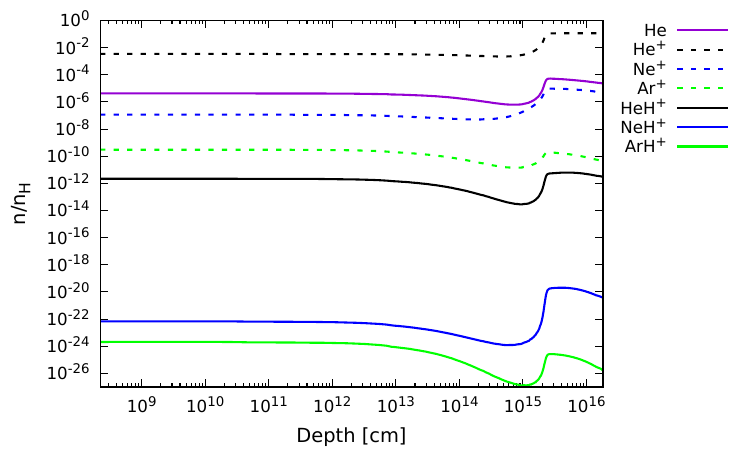}
    \caption{Abundance (with respect to total hydrogen nuclei) profile in QU Vul.}
    \label{fig:abn_qu_vul}
\end{figure}

\autoref{fig:abn_qu_vul} shows the abundance (with respect to total hydrogen nuclei) profile of noble gas hydride cations. HeH$^+$ shows the highest peak abundance ($\sim6.4\times10^{-12}$ with a total column density of {$\sim3.02\times10^9$~cm$^{-2}$}), whereas, ArH$^+$ and NeH$^+$ show negligible abundance ({$2.1\times10^{-24}$ and $2.2\times10^{-20}$}).
ArH$^+$ is mainly formed by
$\rm{Ar^+ + H_2 \rightarrow ArH^+ + H}$, whereas, for HeH$^+$, the main formation pathway is
$\rm{He^+ + H \rightarrow HeH^+ + h\nu}$.
As NeH$^+$ formation by $\rm{Ne^+ + H_2 \rightarrow NeH^+ + H}$ was questioned by \cite{thei15}, we do not consider this reaction in our network. Rather, we notice a favorable formation of NeH$^+$ by $\rm{HeH^+ + Ne \rightarrow NeH^+ + He}$.
As molecular hydrogen is under-abundant in this region, the abundance of ArH$^+$ is insignificant. However, the abundance of HeH$^+$ is favorable because of the comparatively high abundance of atomic hydrogen.
Since NeH$^+$ formation depends on the HeH$^+$ formation, it requires an additional reaction with Ne with a rate constant of $\sim10^{-9}$~cm$^3$~s$^{-1}$. Additionally, the formation of NeH$^+$ is hampered due to the low abundance of Ne.
Another reason for the higher production of HeH$^+$ is that the initial elemental abundance for He is much higher (0.116) than that of Ar ($2.55 \times 10^{-3}$) and Ne ($1.11 \times 10^{-6}$) \citep{schw02}.
The oscillation of the abundances at large depths, as seen in Fig.~\ref{fig:abn_qu_vul}, may be due to the decrease in both hydrogen and electron densities at those depths, as shown in Fig. \ref{fig:phy_qu_vul}.

\begin{figure}
    \centering
    \includegraphics[width=0.49\textwidth]{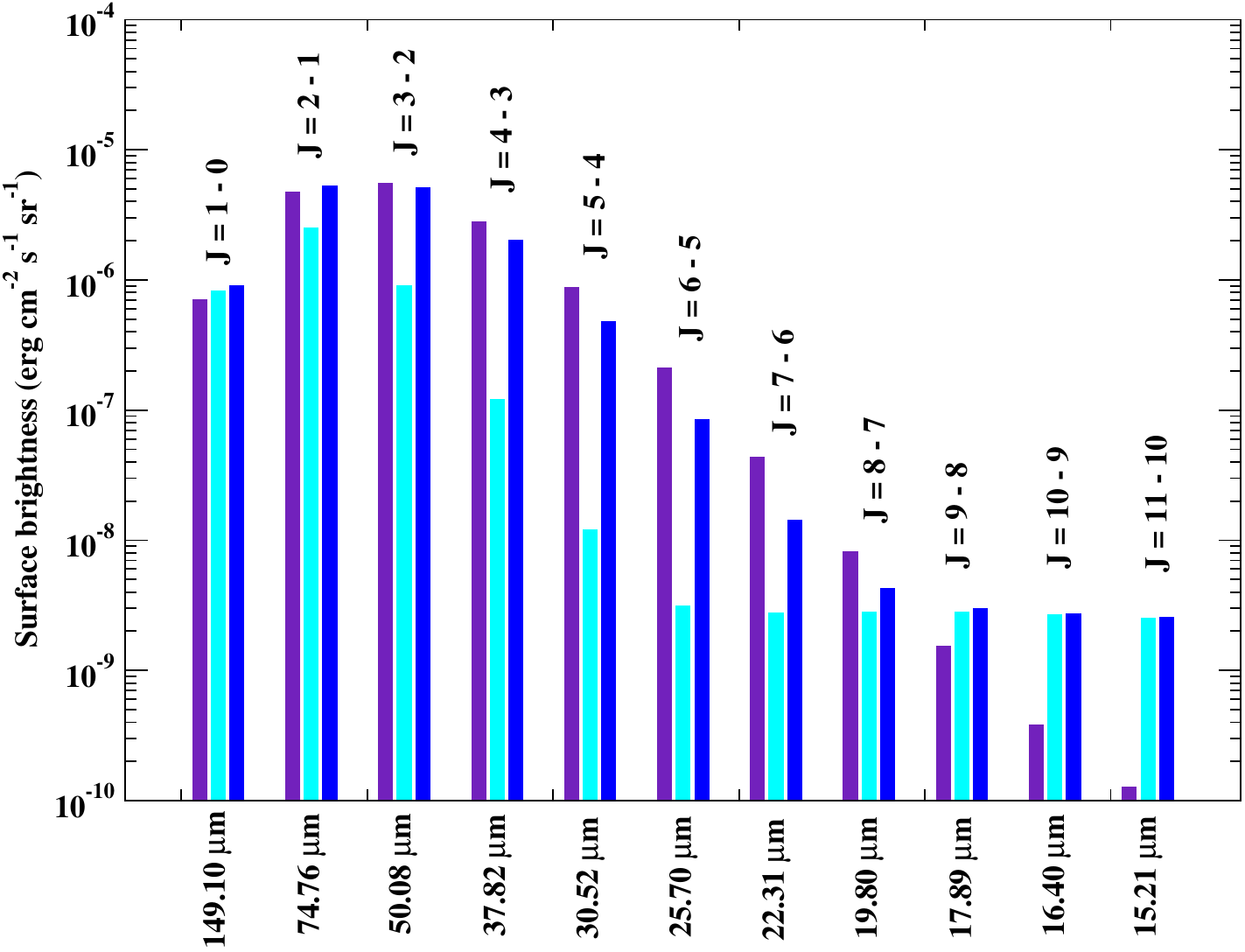}
    \caption{Surface brightness of some of the potential pure rotational transitions of HeH$^+$ in QU Vul by considering the collisional data, which is based on EMAA data (indigo), CB data considering only $\Delta \upsilon=0, \pm 1, \Delta J=\pm1$ (cyan), and CB data considering $\Delta \upsilon=0, \pm 1$, $\Delta J=\pm1$, and $\Delta \upsilon=0$, $\Delta J=\pm2$ (blue). \autoref{tab:HeH+_transitions_qu_vul} summarizes the plotted CB data considering $\Delta \upsilon=0, \pm 1$, $\Delta J=\pm1$, and $\Delta \upsilon=0$, $\Delta J=\pm2$.
    \label{fig:sb_qu_vul}}
\end{figure}

\autoref{fig:sb_qu_vul} shows the intrinsic absolute line surface brightness of the strongest pure rotational transitions of HeH$^+$
in the range 15~--~149~\micron noted in \autoref{tab:HeH+_transitions_qu_vul}.
\autoref{fig:sb_qu_vul} compares surface brightness values to check their dependency on the different collisional data files used for HeH$^+$ (see \autoref{sec:appendix_1}). The indigo color bars show the modeled values considering the EMAA data. The modeled values considering our CB data are shown as the cyan ($\Delta \upsilon=0, \pm 1$, $\Delta J=\pm1$) and blue ($\Delta \upsilon=0, \pm 1$, $\Delta J=\pm1$, and $\Delta \upsilon=0$, $\Delta J=\pm2$) color bars. Figure~\ref{fig:sb_qu_vul}
shows that the impact of collisional data considering $\Delta \upsilon=0$, $\Delta J=\pm2$ (de-)excitation transitions is significant (see \autoref{sec:appendix_1}).
In particular, we note that the new extended data file has little effect for transitions above 17~\micron. In contrast, for higher-frequency transitions ($J_{up} \geq 10$), the surface brightness is significantly increased and is not sensitive to transitions with $\Delta \upsilon=0$, $\Delta J=\pm2$ either.
No effect is seen on the calculated surface brightness values when we consider the EMAA data file with or without hydrogen atoms as colliders, as expected from the low abundance of H. However, a dominant effect is seen for the electron impact collisional rate.
Due to the minimal formation of ArH$^+$ and NeH$^+$, their abundance is very low (see Fig.~\ref{fig:abn_qu_vul}), and hence, we do not calculate the surface brightness of their transitions.

\begin{table*}
\centering
\caption{Pure rotational and ro-vibrational transitions of HeH$^+$ for QU Vul.
\label{tab:HeH+_transitions_qu_vul}}
\begin{tabular}{ccccc}
\hline
{\bf HeH$^+$ lines} & {\bf $\rm{E_U/k_B}$} & {\bf Frequency / wavelength} & {\bf Optical depth} & {\bf Surface brightness} \\
 $(\upsilon^\prime, J^\prime)\to(\upsilon, J)$ & {\bf (K)} & {\bf  (GHz /~\micron)} & {\bf ($\tau$)} & {\bf (erg~cm$^{-2}$~s$^{-1}$~sr$^{-1}$)} \\
\hline
 $(0,1)\to(0,0)$ & 96.48 & 2010.25 / 149.10  & {$1.75\times10^{-5}$} &  {$8.93\times10^{-7}$} \\
 $(0,2)\to(0,1)$ & 288.87 & 4008.87 / 74.76 & {$6.09\times10^{-5}$} & {$5.22\times10^{-6}$}  \\
 $(0,3)\to(0,2)$ & 576.08 & 5984.35 / 50.08 & {$2.50\times10^{-5}$} & {$5.09\times10^{-6}$} \\
 $(0,4)\to(0,3)$ & 956.44 & 7925.43 / 37.82 & {$5.43\times10^{-6}$} & {$2.01\times10^{-6}$}  \\
 $(0,5)\to(0,4)$ & 1427.79 & 9821.24 / 30.52 & {$7.83\times10^{-7}$} & {$4.78\times10^{-7}$}  \\
 $(0,6)\to(0,5)$ & 1987.45 & 11\,661.34 / 25.70 & {$8.90\times10^{-8}$} & {$8.44\times10^{-8}$}  \\
 $(0,7)\to(0,6)$ & 2632.27 & 13\,435.86 / 22.31 & {$8.93\times10^{-9}$} & {$1.43\times10^{-8}$} \\
 $(0,8)\to(0,7)$     & 3358.67 & 15\,135.58 / 19.80 & {$8.56\times10^{-10}$} & {$4.23\times10^{-9}$}  \\
 $(0,9)\to(0,8)$ & 4162.64 & 16\,751.92 / 17.89 & {$9.94\times10^{-11}$} & {$2.96\times10^{-9}$}  \\
 $(0,10)\to(0,9)$ & 5039.81 & 18\,277.00 / 16.40 & {$2.55\times10^{-11}$} & {$2.72\times10^{-9}$}  \\
 $(0,11)\to(0,10)$ &  5985.44 & 19\,703.65 / 15.21 & {$1.38\times10^{-11}$} & {$2.53\times10^{-9}$} \\
\hline
 $(1,0)\to(0,1)\ [P(1)]$ & {4188.26} & {85\,258.39 / 3.51} &  {$1.33\times10^{-6}$} & {$4.25\times10^{-7}$} \\
 $(1,1)\to(0,2)\ [P(2)]$ & {4276.91} & {83\,096.79 / 3.60} & {$6.87\times10^{-7}$} & {$4.66\times10^{-7}$} \\
 $(1,1)\to(0,0)\ [R(0)]$ & {4276.91} & {89\,115.91 / 3.36} & {$1.72\times10^{-6}$} & {$2.61\times10^{-7}$} \\
\hline
\end{tabular}
\end{table*}

The upper-state energy, frequency, and line-center optical depth of the transitions are noted in \autoref{tab:HeH+_transitions_qu_vul}.
From Fig.~\ref{fig:sb_qu_vul} and \autoref{tab:HeH+_transitions_qu_vul}, the most intense transition of HeH$^+$ is the $J=2-1$ at 74.76~\micron (4008~GHz), which shows an absolute intrinsic surface brightness value of {$5.22\times10^{-6}$~erg~cm$^{-2}$~s$^{-1}$~sr$^{-1}$}.
Within the JWST range (0.6~--~28.3~\micron), the maximum surface brightness is obtained for the
$J=6-5$ transition at 25.71~\micron, which shows a surface brightness of {$\sim8.44\times10^{-8}$~erg~cm$^{-2}$~s$^{-1}$~sr$^{-1}$}.


\subsection{RS Oph} \label{sec:rs_oph}
\cite{pand22} constructed a simple phenomenological model assuming a spherical geometry to study the optically thin phase of the 2021 outburst using the photoionization code \textsc{Cloudy}, v.17.02.
RS Oph is a shock-wave generator.
However, the shock velocity it produces decreases gradually over time.
Consequently, a late epoch of day 22 of the 2021 outburst for RS Oph is chosen \citep[see Table 3 of][]{pand22}.

{We use the initial elemental abundances and physical parameters noted in \autoref{tab:phy} for modeling purposes.}
To take all the observed emission lines of both low and high ionization potential into account, we consider two separate density (high-density and low-density) component models with almost the same physical parameters, except $n(H)$ [$n(H)_{dense}=0.35\times10^{10}$~cm$^{-3}$,
$n(H)_{diffuse}=1.77\times10^{8}$~cm$^{-3}$], and with two covering factors (0.61 for dense and 0.39 for diffuse) of the assumed spherically symmetric ejecta, following \cite{pand22}. The resulting geometry is a closed thick shell for both cases.

\begin{figure}
    \centering
    \includegraphics[width=0.49\textwidth]{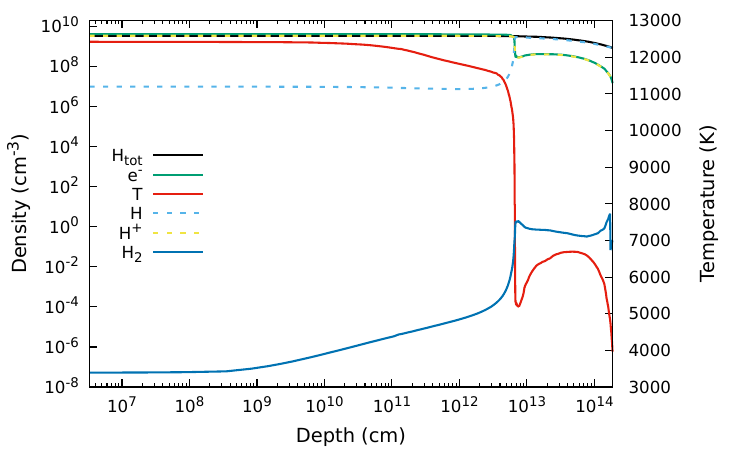}
    \includegraphics[width=0.49\textwidth]{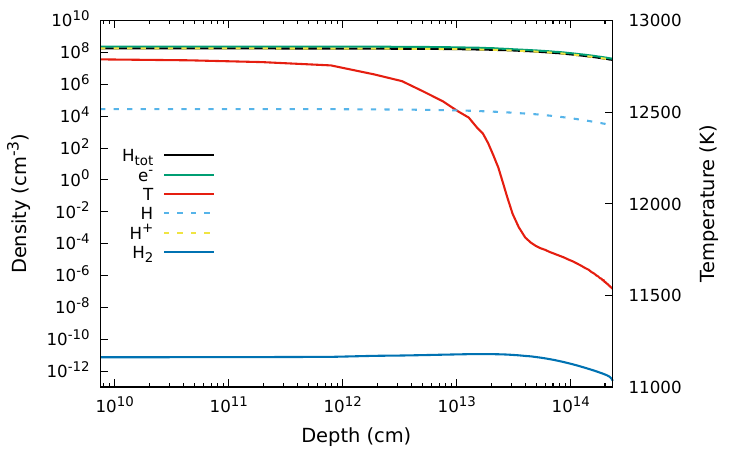}
    \caption{Density and temperature profile in the dense part (upper panel) and diffuse part (lower panel) of the RS Oph ejecta.}
    \label{fig:phy_rs_oph}
\end{figure}

The density variation with depth is shown in Fig.~\ref{fig:phy_rs_oph} for the dense part (upper panel) and diffuse part (lower panel).
The dense part shows a transition zone (at a Str\"{o}mgren radius of $\sim 6.5\times10^{12}$~cm) from ionic (electron and H$^+$) to atomic (H) with an abrupt decrease in temperature and increase in depth, whereas the diffuse part shows a high constant density of electrons and H$^+$ throughout the depth.
The electron temperature varies within ranges of 4000~--~12\,400~K and 11\,500~--~12\,800~K for the dense and diffuse parts, respectively.

\begin{figure}
    \centering
    \includegraphics[width=0.49\textwidth]{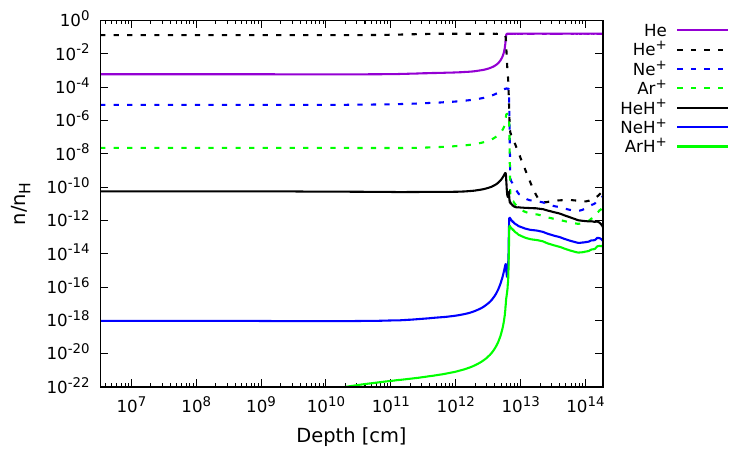}
    \includegraphics[width=0.49\textwidth]{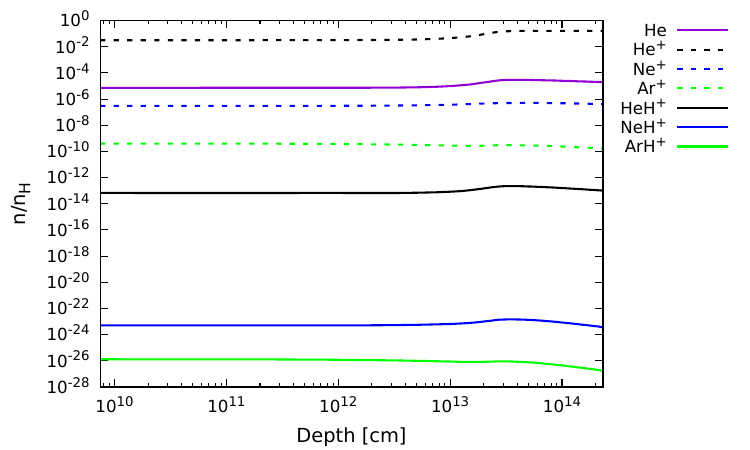}
    \caption{Abundance with respect to total hydrogen in all forms for the dense part (upper panel) and diffuse part (lower panel) in RS Oph.}
    \label{fig:abn_rs_oph}
\end{figure}

In the dense component (see the upper panel of Fig.~\ref{fig:abn_rs_oph}), we have significant production of HeH$^+$ with a peak abundance of $\sim6.88\times10^{-10}$ and a total column density of $\sim4.00\times10^{10}$~cm$^{-2}$. As the abundance of H$_2$ in the dense component of RS Oph is much higher ($\sim5.80\times10^{-9}$) than QU Vul ($\sim3.05\times10^{-17}$), the formation of ArH$^+$ is comparatively favorable. Similarly, a comparatively high abundance of H accelerates the formation of NeH$^+$ from HeH$^+$. At the dense component, we have peak abundances of {$\sim4.47\times10^{-13}$ and $\sim1.40\times10^{-12}$} for ArH$^+$ and NeH$^+$, respectively.

\begin{figure}
    \centering
    \includegraphics[width=0.49\textwidth]{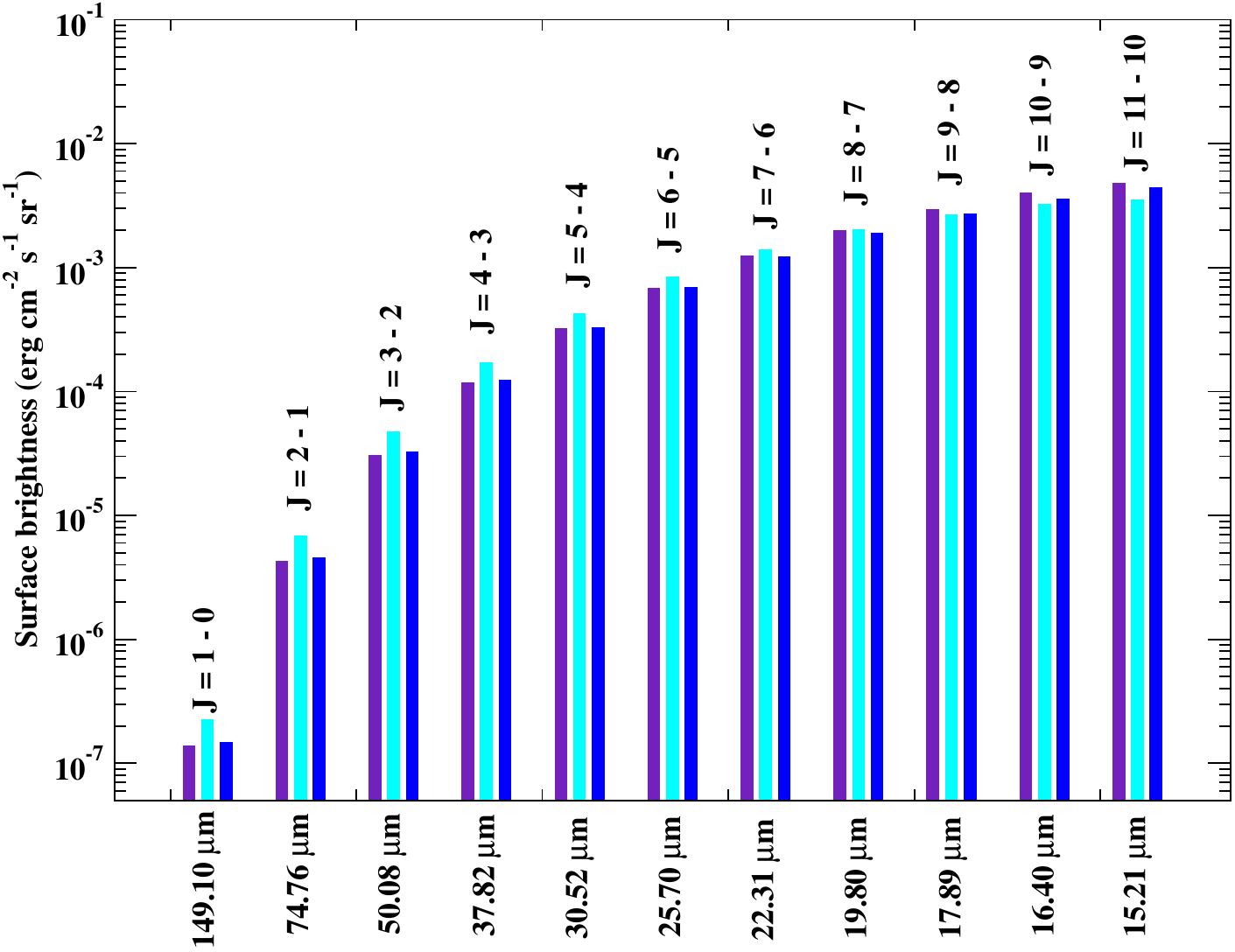}
    \includegraphics[width=0.49\textwidth]{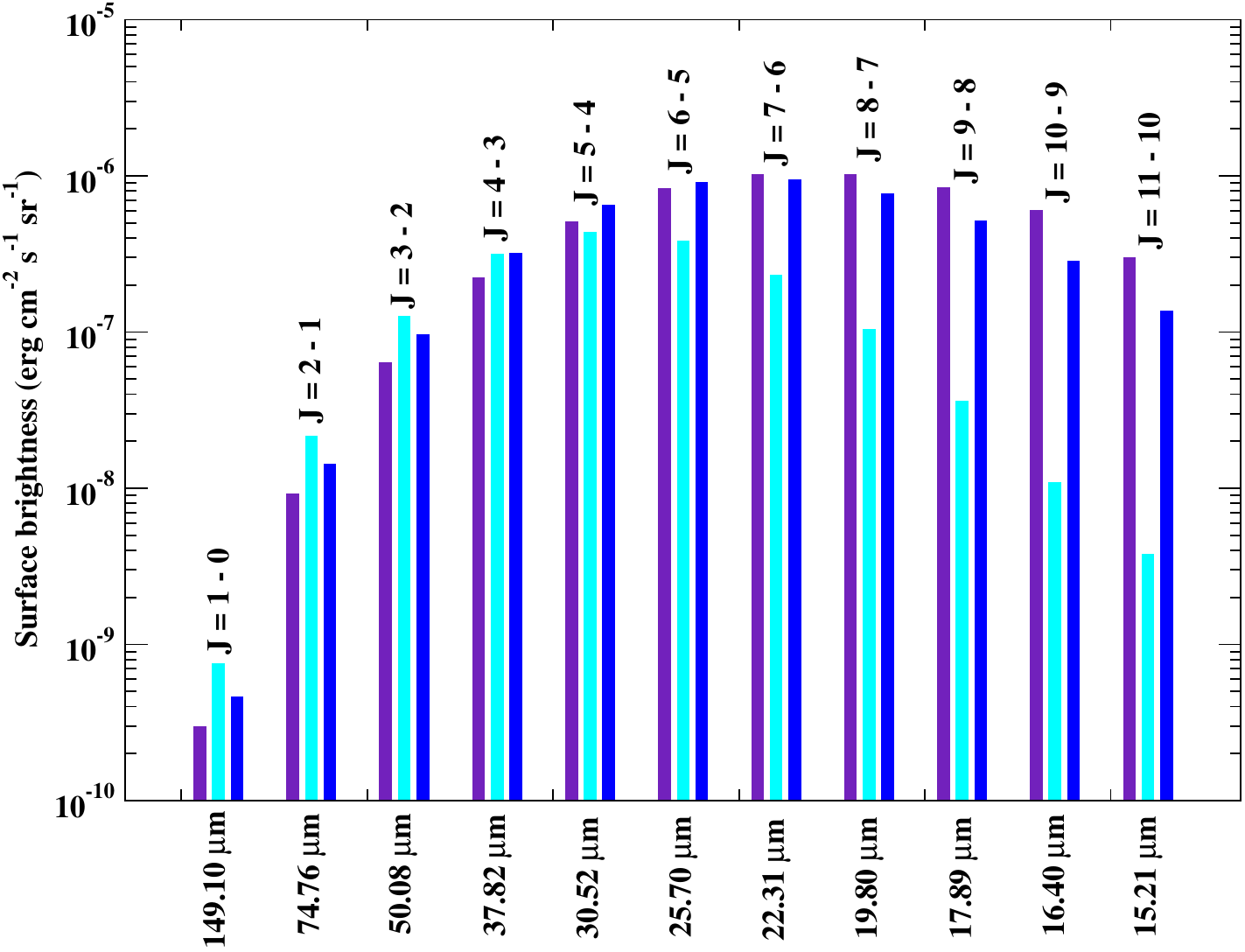}
    \caption{Comparison of the surface brightness of some possible HeH$^+$ pure rotational transitions in the dense (upper panel) and diffuse (lower
panel) parts of RS Oph  calculated using \textsc{Cloudy,} considering HeH$^+$ collisional data based on EMAA data (indigo), CB data considering only $\Delta \upsilon=0, \pm 1$, $\Delta J=\pm1$ (cyan), and CB data considering both $\Delta \upsilon=0, \pm 1$, $\Delta J=\pm1$, and $\Delta \upsilon=0$, $\Delta J=\pm2$ (blue). The plotted CB data considering $\Delta \upsilon=0, \pm 1$, $\Delta J=\pm1$, and $\Delta \upsilon=0$, $\Delta J=\pm2$ are summarized in \autoref{tab:HeH+_transitions_rs_oph}.
    \label{fig:sb_rs_oph_HeH+}}
\end{figure}

\begin{table*} 
\caption{Pure rotational and ro-vibrational transitions of HeH$^+$ for RS Oph.
\label{tab:HeH+_transitions_rs_oph}}
\centering
\resizebox{\linewidth}{!}{\begin{tabular}{ccccc}
\hline
 {\bf HeH$^+$ lines} & {\bf $\rm{E_U/k_B}$} & {\bf Frequency / Wavelength} & {\bf Optical Depth ($\tau$)} & {\bf Surface Brightness (erg~cm$^{-2}$~s$^{-1}$~sr$^{-1}$)} \\ 
  $(\upsilon^\prime, J^\prime)\to(\upsilon, J)$ & {\bf (K)} & {\bf  (GHz /~\micron)} & {\bf Dense / Diffuse}  & {\bf Dense / Diffuse} \\
\hline
 $(0,1)\to(0,0)$ & 96.48 & 2010.25 / 149.10 & {$3.51\times10^{-7}$ / $3.79\times10^{-10}$} & {$1.47\times10^{-7}$} / $4.58\times10^{-10}$ \\
 $(0,2)\to(0,1)$ & 288.87 & 4008.87 / 74.76 & {$1.59\times10^{-6}$ / $3.16\times10^{-9}$} & {$4.53\times10^{-6}$} / $1.41\times10^{-8}$ \\
 $(0,3)\to(0,2)$ & 576.08 & 5984.35 / 50.08 & {$4.24\times10^{-6}$ / $1.38\times10^{-8}$} & {$3.23\times10^{-5}$} / $9.58\times10^{-8}$ \\
 $(0,4)\to(0,3)$ & 956.44 & 7925.43 / 37.82  & {$8.50\times10^{-6}$ / $3.57\times10^{-8}$} & {$1.23\times10^{-4}$} / $3.18\times10^{-7}$ \\
 $(0,5)\to(0,4)$ & 1427.79 & 9821.24 / 30.52 & {$1.37\times10^{-5}$ / $5.95\times10^{-8}$} & {$3.26\times10^{-4}$} / $6.47\times10^{-7}$ \\
 $(0,6)\to(0,5)$ & 1987.45 & 11\,661.34 / 25.70 & {$1.87\times10^{-5}$ / $6.93\times10^{-8}$} &  {$6.84\times10^{-4}$} / $9.08\times10^{-7}$ \\
 $(0,7)\to(0,6)$ & 2632.27 & 13\,435.86 / 22.31 & {$2.26\times10^{-5}$ / $6.04\times10^{-8}$} &  {$1.21\times10^{-3}$} / $9.44\times10^{-7}$ \\
 $(0,8)\to(0,7)$ & 3358.67 & 15\,135.58 / 19.80 & {$2.49\times10^{-5}$ / $4.18\times10^{-8}$} &  {$1.89\times10^{-3}$} / $7.68\times10^{-7}$ \\
 $(0,9)\to(0,8)$ & 4162.64 & 16\,751.92 / 17.89 & {$2.56\times10^{-5}$ / $2.39\times10^{-8}$} &  {$2.69\times10^{-3}$} / $5.09\times10^{-7}$ \\
 $(0,10)\to(0,9)$ & 5039.81 & 18\,277.00 / 16.40 & {$2.50\times10^{-5}$ / $1.16\times10^{-8}$} &  {$3.54\times10^{-3}$} / $2.83\times10^{-7}$ \\
 $(0,11)\to(0,10)$ &  5985.44 & 19\,703.65 / 15.21 & {$2.38\times10^{-5}$ / $4.89\times10^{-9}$} & {$4.36\times10^{-3}$} / $1.36\times10^{-7}$ \\
\hline
 $(1,0)\to(0,1)\ [P(1)]$ & {4188.26} & {85\,258.39 / 3.51} &  {$9.53\times10^{-7}$/ $1.79\times10^{-9}$} & {$2.75\times10^{-4}$} / $3.57\times10^{-8}$ \\
 $(1,1)\to(0,2)\ [P(2)]$ & {4276.91} & {83\,096.79 / 3.60} & {$6.86\times10^{-7}$ / $3.67\times10^{-9}$} & {$5.23\times10^{-4}$}  / $6.77\times10^{-8}$ \\
 $(1,1)\to(0,0)\ [R(0)]$ & {4276.91} & {89\,115.91 / 3.36} & {$8.66\times10^{-7}$ / $1.62\times10^{-9}$} & {$2.93\times10^{-4}$}  / $3.80\times10^{-8}$ \\
\hline
\end{tabular}}
\end{table*}

\autoref{fig:sb_rs_oph_HeH+} shows the predicted absolute intrinsic line surface brightness (values are noted in \autoref{tab:HeH+_transitions_rs_oph}) of the strongest pure rotational transitions of HeH$^+$ in the 15~--~149~\micron range.
The most intense transition of HeH$^+$ is predicted to be $J=11-10$ at 15.21~\micron (19\,703~GHz) for the dense part and $J=7-6$ at 22.31~\micron (13\,435~GHz) for the diffuse part.
Similar to Fig.~\ref{fig:sb_qu_vul}, a comparison has been conducted by considering different collisional data files of HeH$^+$. 
In the dense component, the electron density is larger than the density of the hydrogen atom, where HeH$^+$ abundance reached the peak (see the upper panels of Figs.~\ref{fig:phy_rs_oph} and \ref{fig:abn_rs_oph}). However, to see the impact of hydrogen atoms as collision partners, we check surface brightness values. We consider both electrons and hydrogen atoms as collision partners with the EMAA data file and only electrons with the CB data file. We do not see any significant impact, which is due to the much larger rate coefficients for electrons ($\sim10^{-6}$~cm$^3$~s$^{-1}$) than for H ($\sim10^{-10}$~cm$^3$~s$^{-1}$).
\autoref{tab:HeH+_transitions_rs_oph} depicts the calculated surface brightness by considering the HeH$^+$ collisional data file obtained with CB approximation with $\Delta \upsilon=0, \pm 1$, $\Delta J=\pm1$, and $\Delta \upsilon=0$, $\Delta J=\pm2$ (see \autoref{sec:appendix_1}).
It is clear from both the panels of Fig.~\ref{fig:sb_rs_oph_HeH+} and \autoref{tab:HeH+_transitions_rs_oph} that the dense part produces brighter transitions than the diffuse part of the ejecta.
 
From the upper panel of Fig. \ref{fig:sb_rs_oph_HeH+} and \autoref{tab:HeH+_transitions_rs_oph}, the strongest transition of HeH$^+$ is the $J=11-10$ at 15.21~\micron (19\,703~GHz), which also lies within the JWST range, considering the dense part of the RS Oph ejecta, which shows an absolute intrinsic surface brightness value of $4.36\times10^{-3}$~erg~cm$^{-2}$~s$^{-1}$~sr$^{-1}$.
We also calculate the surface brightness of pure rotational transitions of {$\rm{^{36}ArH^+}$} and NeH$^+$ for the dense part of the ejecta (see \autoref{sec:appendix_2}). However, as previously stated, the lack of high-temperature data could result in inaccurate estimations.

\subsection{V1716 Sco} \label{sec:v1716_sco}
We consider another CNe, V1716 Sco, as a third case study.
This choice is based on the most recent work by \cite{wood24} and their prediction of HeH$^+$ production in an observable amount in the denser component of V1716 Sco.
Therefore, we only present the analysis of the high-density clump component of V1716 Sco in the coronal line phase of its evolution 132.8 days after outburst \citep{wood23}, focusing on the proposed best-fit physical parameters and initial elemental abundances, as detailed in \autoref{tab:phy}.
A cylindrical (truncated spherical) geometry of gas expanding with a velocity of $\rm{v_{exp} = 1000}$~km~s$^{-1}$ was assumed by \cite{wood24} to model the infrared spectroscopic data of V1716 Sco.
The resulting geometry is a closed thick shell.

\begin{figure}
    \centering
    \includegraphics[width=0.5\textwidth]{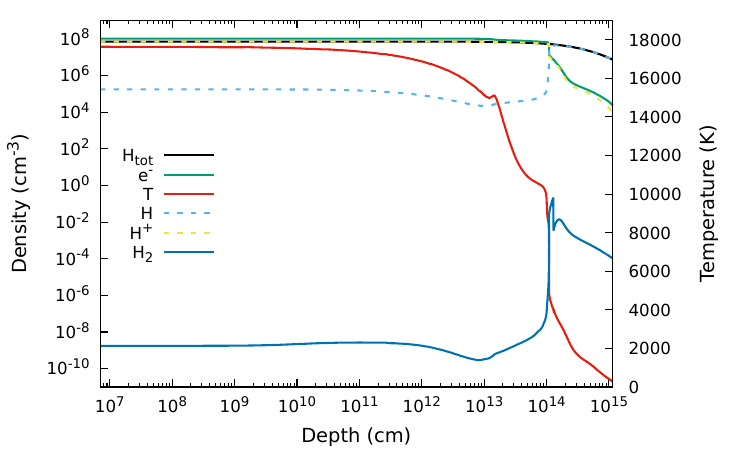}
    \caption{Density and temperature profile in the dense part of V1716 Sco ejecta.}
    \label{fig:phy_v1716_sco}
\end{figure}

\begin{figure}
    \centering
    \includegraphics[width=0.50\textwidth]{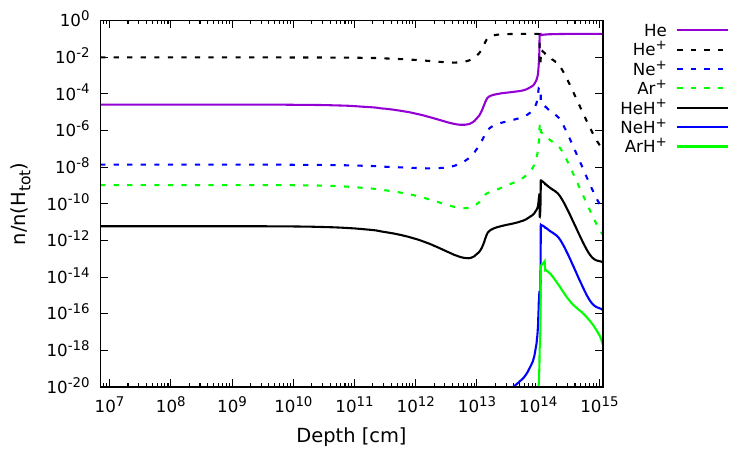}
    \caption{Abundance (with respect to total hydrogen nuclei) profile in the dense part of V1716 Sco ejecta.}
    \label{fig:abn_v1716_sco}
\end{figure}

The density variation with depth is depicted in Fig.~\ref{fig:phy_v1716_sco} for the dense part of the ejecta.
The gas is highly ionized and becomes neutral atomic toward the outer edge at a Str\"{o}mgren radius of $\sim1.10\times10^{14}$~cm.
The electron temperature decreases from 17\,638~K to 292~K with increasing distance from the source.
The abundance (with respect to total hydrogen nuclei) profiles of all noble gas hydride ions, along with their atomic ions and He, are shown in Fig~\ref{fig:abn_v1716_sco}.
As noted by \cite{wood24}, we also find an observable amount of HeH$^+$ with a peak abundance of $\sim 1.88 \times 10^{-9}$ near the overlapping Str\"{o}mgren spheres at a substantial column density of {$\sim9.60\times10^{11}$~cm$^{-2}$} considering the chemical reaction set noted in \autoref{table:reaction_Ar}, with their corresponding rate coefficients.
\cite{wood24} obtained a HeH$^+$ column density of $\sim9.10\times10^{12}$~cm$^{-2}$ from their modeling with an almost one-order-of-magnitude higher abundance than ours.
The lower HeH$^+$ column density is due to the modified rate coefficients of the dominating formation and destruction reactions of HeH$^+$ \citep{neuf20} considered in this study (see \autoref{table:reaction_Ar}).
We obtain peak abundances of {$\sim7.18\times10^{-14}$ and $\sim 6.82\times10^{-12}$} of ArH$^+$ and NeH$^+$, respectively.

\begin{figure}
    \centering
    \includegraphics[width=0.49\textwidth]{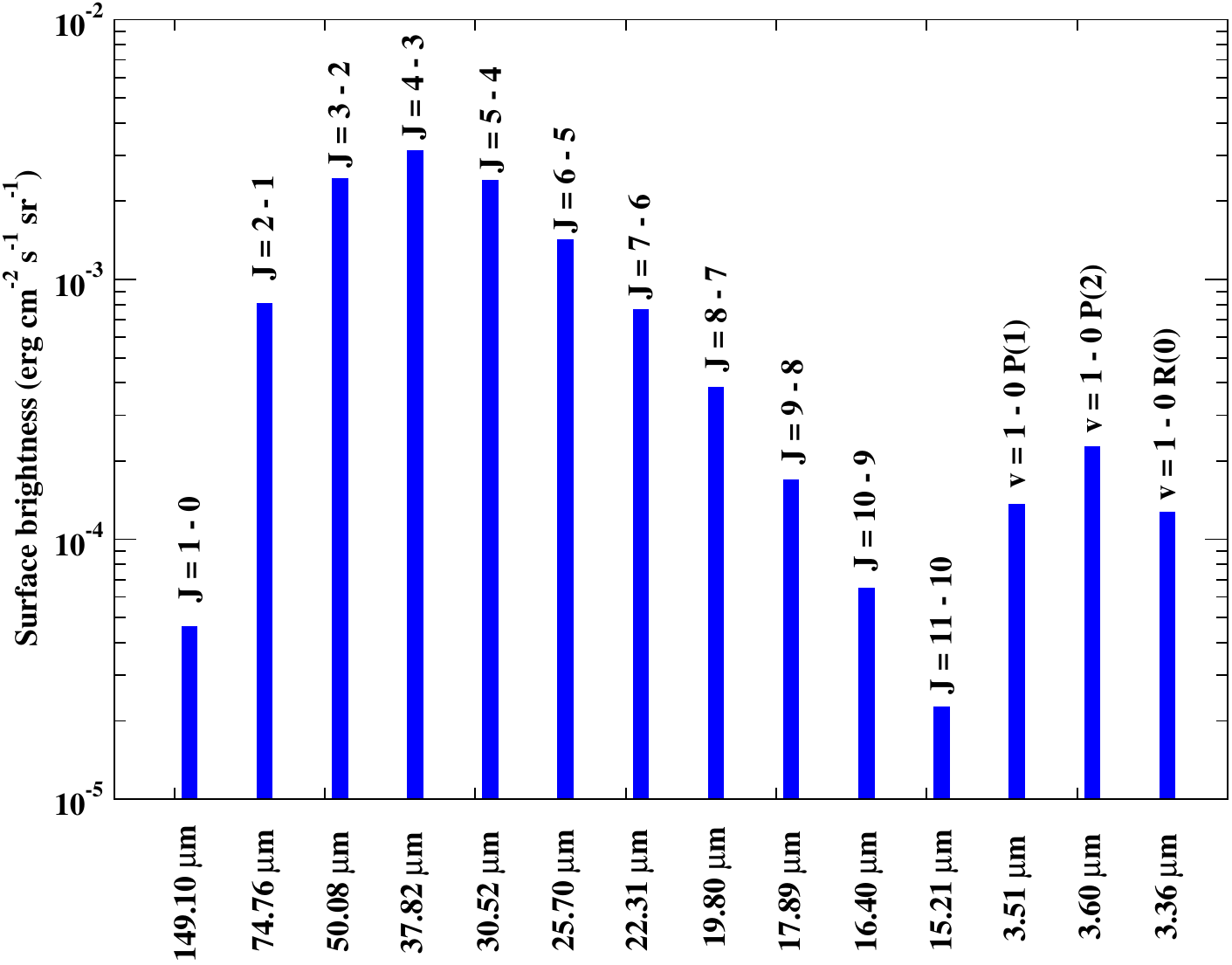}
    \caption{Surface brightness of some of the potential pure rotational and ro-vibrational transitions of HeH$^+$ in V1716 Sco by considering the collisional data based on CB data considering $\Delta \upsilon=0, \pm 1$, $\Delta J=\pm1$, and $\Delta \upsilon=0$, $\Delta J=\pm2$ (blue). \autoref{tab:HeH+_transitions_v1716_sco} summarizes the plotted CB data considering $\Delta \upsilon=0, \pm 1$, $\Delta J=\pm1$, and $\Delta \upsilon=0$, $\Delta J=\pm2$.
    \label{fig:sb_heh+_v1716_sco}}
\end{figure}

\autoref{fig:sb_heh+_v1716_sco} shows the predicted (based on HeH$^+$ collisional data file obtained with the CB approximation with $\Delta \upsilon=0, \pm 1$, $\Delta J=\pm1$, and $\Delta \upsilon=0$, $\Delta J=\pm2$ (see \autoref{sec:appendix_1})) absolute intrinsic line surface brightness (values are noted in \autoref{tab:HeH+_transitions_v1716_sco}) of the strongest pure rotational transitions of HeH$^+$ and the three most intense ro-vibrational transitions.
The most intense transition of HeH$^+$ is predicted to be $J=4-3$ at 37.82~\micron (7925~GHz).
However, within the JWST range, the strongest transition of HeH$^+$ is predicted to be $J=6-5$ at 25.70~\micron (11\,661~GHz).
Based on the \textsc{Cloudy} number densities, we also calculated the line surface brightnesses of the three ro-vibrational transitions, $\upsilon = 1-0\ P(1),\ P(2)$, and $R(0)$ to decipher whether or not they are likely to be detectable, which is further discussed in Sect.~\ref{sec:jwst}.
We also calculated the surface brightness of pure rotational transitions of {$\rm{^{36}ArH^+}$} and NeH$^+$ for the dense part of the ejecta and provide these in \autoref{sec:appendix_2}. However, as mentioned above, the absence of high-temperature data could lead to inaccurate estimates.

\begin{table*}
\caption{Pure rotational and ro-vibrational transitions of HeH$^+$ for V1716 Sco.
\label{tab:HeH+_transitions_v1716_sco}}
\centering
\begin{tabular}{ccccc}
\hline
 {\bf HeH$^+$ lines} & {\bf $\rm{E_U/k_B}$} & {\bf Frequency / wavelength} & {\bf Optical depth} & {\bf Surface brightness} \\
 $(\upsilon^\prime, J^\prime)\to(\upsilon, J)$ & {\bf (K)} & {\bf  (GHz /~\micron)} & {\bf ($\tau$)} & {\bf (erg~cm$^{-2}$~s$^{-1}$~sr$^{-1}$)} \\
\hline
 $(0,1)\to(0,0)$ & 96.48 & 2010.25 / 149.10 & {$8.31\times10^{-4}$} & {$4.51\times10^{-5}$}  \\
 $(0,2)\to(0,1)$ & 288.87 & 4008.87 / 74.76 & {$1.28\times10^{-2}$} & {$7.91\times10^{-4}$}  \\
 $(0,3)\to(0,2)$ & 576.08 & 5984.35 / 50.08 & {$1.35\times10^{-2}$} & {$2.38\times10^{-3}$} \\
 $(0,4)\to(0,3)$ & 956.44 & 7925.43 / 37.82 & {$8.02\times10^{-3}$} & {$3.05\times10^{-3}$} \\
 $(0,5)\to(0,4)$ & 1427.79 & 9821.24 / 30.52  & {$3.25\times10^{-3}$} & {$2.35\times10^{-3}$} \\
 $(0,6)\to(0,5)$ & 1987.45 & 11\,661.34 / 25.70  & {$9.88\times10^{-4}$} &  {$1.39\times10^{-3}$}  \\
 $(0,7)\to(0,6)$ & 2632.27 & 13\,435.86 / 22.31   & {$2.60\times10^{-4}$} & {$7.50\times10^{-4}$} \\
 $(0,8)\to(0,7)$ & 3358.67 & 15\,135.58 / 19.80  & {$6.95\times10^{-5}$} & {$3.76\times10^{-4}$}  \\
 $(0,9)\to(0,8)$ & 4162.64 & 16\,751.92 / 17.89  & {$2.00\times10^{-5}$} & {$1.66\times10^{-4}$} \\
 $(0,10)\to(0,9)$ & 5039.81 & 18\,277.00 / 16.40  & {$5.76\times10^{-6}$} & {$6.34\times10^{-5}$} \\
 $(0,11)\to(0,10)$ &  5985.44 & 19\,703.65 / 15.21  & {$1.55\times10^{-6}$} & {$2.18\times10^{-5}$} \\
\hline
 $(1,0)\to(0,1)\ [P(1)]$ &  4188.26 & 85\,258.39 / 3.51 &  {$4.52\times10^{-4}$} & {$1.33\times10^{-4}$} \\
 $(1,1)\to(0,2)\ [P(2)]$ &  4276.91 & 83\,096.79 / 3.60 & {$5.08\times10^{-4}$} & {$2.21\times10^{-4}$} \\
 $(1,1)\to(0,0)\ [R(0)]$ &  4276.91 & 89\,115.91 / 3.36 & {$4.30\times10^{-4}$} & {$1.24\times10^{-4}$} \\
\hline
\end{tabular}
\end{table*}

\begin{figure}
    \centering
    \includegraphics[width=0.49\textwidth]{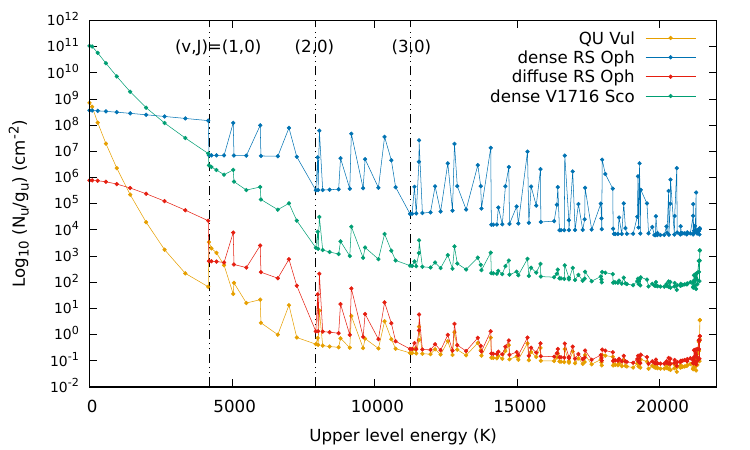}
   \caption{ HeH$^+$~($\upsilon, J$) population diagram for QU Vul (orange), RS Oph dense part (blue), RS Oph diffuse part (red), and V1716 Sco dense part (green).
    \label{fig:heh+_population_diagram}}
\end{figure}

Finally, Fig.~\ref{fig:heh+_population_diagram} illustrates the relative population per sub-level as a function of the upper-level energy for QU Vul (orange), the dense part of RS Oph (blue), the diffuse part of RS Oph (red), and the dense part of V1716 Sco (green).
The relative population per sub-level of all the {162} ro-vibrational energy levels, up to the highest level {$\upsilon = 11,\ J = 1,$ which lies about 21\,400~K} above the ground level, are plotted considering the HeH$^+$ collisional data file obtained with the CB approximation with $\Delta \upsilon=0, \pm 1$, $\Delta J=\pm1$, and $\Delta \upsilon=0$, $\Delta J=\pm2$ (see \autoref{sec:appendix_1}).

\subsection{HeH$^+$ formation timescale}
For this modeling, we assume the gas is in steady-state equilibrium.
We understand that this assumption is most likely invalid because of the rapidly changing dynamic nature of the nova outburst region, yet we wish to check the microphysical timescales involved in this study to verify whether our simplistic \textsc{Cloudy} model is time-steady.

For QU Vul, we obtain the longest thermal timescale of $\sim7.81 \times 10^6$~s ($\sim 90$~days), a 21~cm equilibrium timescale of $\sim2.30 \times 10^3$~s ($\sim38$~minutes), and the longest H recombination timescale of $\sim6.05 \times 10^7$~s ($\sim$1.92~years~$\sim702$~days). As the longest calculated timescale ($\sim$1.92~years~$\sim702$~days) is comparable to the age of the cloud (day 550 $\sim1.51$~years), our estimated surface brightness should be considered the upper limit of the transition.

In the case of the dense part of the RS Oph ejecta model, we obtain $\sim1.27 \times 10^6$~s ($14.7$~days), $1.93$~s, and $1.33 \times 10^5$~s (1.5~days) for the longest thermal timescale, a 21~cm equilibrium timescale, and the longest H-recombination timescale, respectively.
Therefore, the longest timescale is  $14.7$~days, and we are considering the 22$^{nd}$~day ($1.9\times10^6$~s) after the outburst occurs.
For the diffuse part, we obtain an even lower longest timescale of 1.34 days, and so in these cases our calculated surface brightness is well constrained within the age of the cloud, and the system is in photoionization equilibrium.
The longest time recorded is 14.5 days for the dense part, meaning the parameters should remain constant over this period. However, the situation is more dynamic in earlier days. Therefore, it is important to note that the value obtained here should be considered an upper limit.

In the case of the V1716 Sco dense component model, the longest thermal timescale is found to be $\sim 970$~days, or $\sim2.66$~years.
This suggests that the equilibrium timescale for the thermal solution is longer than the age of the gas cloud.
However, \cite{wood24} provided the best-fit model for the epoch day 132.8, which is shorter than the longest thermal timescale based on their best-fit \textsc{Cloudy} model, and atomic processes or microphysics cannot become time-steady within this period.

\subsection{Detectability} \label{sec:jwst}
The atmospheric transmission of Earth limits the ground-based detection of the forecasted HeH$^+$ transitions from suitable sites (e.g., Cerro Chajnantor, Chile) to a selection of spectral lines. For a water-vapor column of 0.2~mm (pwv), the $J=11-10$ and $J=10-9$ lines are blocked and hampered by the atmosphere, respectively. The transitions $J=9-8$ to $J=6-5$ are possible for high-resolution spectroscopy ($R \sim10^6$, so as to correct for individual, narrow atmospheric features). The rotational transitions between lower $J$ levels are reserved for airborne or space-based observatories. Currently, only the MIRI/MRS instrument on board the JWST provides access to the modeled mid-infrared observations, at moderate spectral resolution \citep[$R = \lambda/\Delta\lambda \sim$1000~--~3000,][]{argyriou23}. In the following, we assess their possibility to detect.

At a distance of 2.68~kpc \citep{schaefer22} and for the modeled transitions accessible to MIRI/MRS ($J=6-5$ to $J=11-10$), the line-area-integrated flux densities emitted from the dense component of RS Oph amount to $(0.75-5.71)\times 10^{-17}$~erg~cm$^{-2}$~s$^{-1}$. Following the JWST exposure time calculator \citep[version 3.2,][]{pontoppidan16}, with a total exposure time of approximately four hours, the $J=6-5$ line ($\lambda\,25.70$~\micron) will be undetectable, with a signal-to-noise ratio (S/N) of 0.07, while the $J=11-10$ line ($\lambda\,15.21$~\micron) reaches a detectable S/N\ of 8.3. These estimates are for spectroscopy with the integral field unit, two-point nodding, and sensitivity-optimizing apertures of 0.6\arcsec \ and 0.3\arcsec, respectively, accounting for the sky background brightness. We note that the $J=10-9$ line ($\lambda\,16.40$~\micron) of HeH$^+$ is also detectable with the same setup, at a S/N of 6.4. However, these rather promising evaluations do not account for the possibility of spectral blends with infrared recombination lines (e.g., \ion{H}{i} (21-11) at $\lambda\,15.20$~\micron and \ion{H}{i} (15-10) at $\lambda\,16.41$~\micron); avoiding these would require a spectral resolution of $R \ga 1500$.
For V1716 Sco, the feasibility estimate for mid-infrared spectroscopy is less optimistic, although its larger size compensates for the greater distance \citep[3.6~kpc;][]{wood24}.
With the same technical and observational setup as for RS Oph, the $J=6-5$ line would appear with 1.1~$\sigma_\mathrm{rms}$, while the $J=11-10$ line displays only 0.3~$\sigma_\mathrm{rms}$.
Even for a marginal 3~$\sigma_\mathrm{rms}$ detection, $\sim 30$~hours telescope time would be required.

For the ro-vibrational $\upsilon=1-0$ transitions in the near-infrared, the situation is more promising: Using the Near Infrared Spectrograph (NIRSpec) instrument on board  JWST \citep{jako22} in a high-spectral-resolution ($R\sim 2700$) single-slit configuration, after one hour integration time the $P(1)$, $P(2),$ and $R(0)$ lines of the dense component of V1716 Sco at $\lambda~3.515$, 3.607, and 3.363~\micron would be detected at, respectively, 5.6, 8.8, and 5.1~$\sigma_\mathrm{rms}$. For the dense part of RS Oph, the corresponding results are, respectively, 2.0, 1.9, and 3.5~$\sigma_\mathrm{rms}$, justifying a longer but reasonable integration time of a few hours. We note, however, that the near-infrared continuum brightness of V1716 Sco, as reported by \cite{wood24}, bears the risk of detector saturation, at least partially.

However, with this study of the detectability of HeH$^+$ transitions in these sources, our aim is to assess the potential for their future detection in objects with comparable physical and chemical conditions.
Monitoring the recurrent occurrences of novae in the nearby vicinity presents a unique opportunity to track HeH$^+$ in real time, enabling us to observe their evolution as the nova progresses.
We can gather evidence that either supports the presence of HeH$^+$ or offers definitive data that can be used to challenge or rule out current hypotheses.


\section{Conclusions} \label{sec:conclusions}

HeH$^+$ was the first heteronuclear molecule formed in the metal-free Universe after the Big Bang. Its presence has long been speculated in interstellar space.
Although it was first identified in the laboratory nearly a century ago \citep{hogn25}, the possibility that HeH$^+$ might exist in local astrophysical plasmas, especially in the planetary nebula, was proposed in the late 1970s \citep{blac78}.
After its first circumstellar detection in the young and dense planetary nebula, NGC 7027 \citep{gust19,neuf20}, the molecule received particular attention.
Here, we report the possibility of identifying HeH$^+$ {in similar physical conditions to CNe ejecta} with space-based observation (specifically by the JWST).
We carry out a photoionization modeling {(based on the earlier published best-fit physical parameters)} for the nova outburst region to study the fate of noble gas-related species.
Our findings are summarized as follows:

   \begin{enumerate}

      \item We find several pure rotational and ro-vibrational transitions of HeH$^+$ that fall in the range of JWST and  have observable surface brightness in both the dense component of RS Oph and V1716 Sco nova ejecta regions. The surface brightness values are larger in RS Oph and V1716 Sco than in QU Vul due to the much larger electron density and, to a lesser extent, the larger HeH$^+$ abundance. Here, we consider steady-state calculations and check the microphysical timescales, which suggest that the surface brightnesses that we have computed for these transitions should be considered as upper limits.
      Nevertheless, our feasibility estimate for the JWST predicts that toward RS~Oph, a S/N of up to 8 can be reached in the $J=11-10$ line after four hours of integration time using the MIRI/MST instrument, leaving a sufficient margin for a potentially significant detection.
      For the ro-vibrational $\upsilon = 1-0,$ $P(1)$, $P(2)$, and $R(0)$ transitions in the near-infrared toward V1716 Sco would have been detected at a S/N of 5.6, 8.8, and 5.1, respectively, using the JWST NIRSpec instrument after one hour of integration time. For RS Oph, the corresponding results suggest a longer but reasonable integration time of a few hours. However, there is a risk of detector saturation in the near-infrared continuum brightness of V1716 Sco.
     It should be noted that the CNe objects studied here are used as templates, and not as targets for observations.
      The detection of these lines toward similar objects could be important for deriving the physical conditions of these systems and for testing our chemical model prediction based on our chemical network and new electron-impact ro-vibrational collisional data for HeH$^+$ at temperatures of up to 20\,000~K.
      Furthermore, the application of our chemistry network and collisional data is not limited to novae and can be extended to other settings as well.

      \item Due to the lower abundance of molecular hydrogen (specifically in QU Vul), the abundance of ArH$^+$ is negligible. As NeH$^+$ in our reaction network depends on the formation of HeH$^+$, its formation is slowed down. However, we obtain a high abundance of HeH$^+$ because of the high abundance of atomic hydrogen.

      \item We cannot estimate the surface brightness of the transitions of {$\rm{^{36}ArH^+}$} and NeH$^+$ falling within the JWST domain because of the unavailability of the collisional de-excitation rates for a sufficient number of energy levels. However, we note that in the denser parts of RS Oph and V1716 Sco, an intense surface brightness can be expected for {$\rm{^{36}ArH^+}$} and NeH$^+$ transitions due to a very high abundance of electrons.

    \end{enumerate}

Finally, we note that HeH$^+$ belongs to the class of ``reactive'' ions that can be destroyed so quickly that their level populations never get fully equilibrated, as observed, for example, in CH$^+$ \citep[][and references therein]{Faure2017} and OH \citep{zann24}. For such species, chemical formation and destruction rates are in competition with inelastic rates and should be included in the statistical equilibrium equations. This will be investigated in future work.

\section*{Data availability}
The data underlying this article are made available under a Creative Commons Attribution license on Zenodo: doi:\href{https://doi.org/10.5281/zenodo.14097180}{10.5281/zenodo.14097180}.

\begin{acknowledgements}
      {The authors thank the referees for the valuable insights and constructive critiques that significantly improved the manuscript.}
      M.S. acknowledges financial support from the European Research Council (consolidated grant COLLEXISM, grant agreement ID: 811363). M.S. and R.D. would like to acknowledge S. N. Bose National Centre for Basic Sciences, Salt Lake, Kolkata, under the Department of Science and Technology (DST), Government of India. A.D. acknowledges the Max Planck Society for sponsoring a scientific visit of three months. R.P. acknowledges the Physical Research Laboratory, Department of Space, Government of India, for her Post-doctorate fellowship. P.C. acknowledges the support of the Max Planck Society. This research has made use of spectroscopic and collisional data from the EMAA database (\url{https://emaa.osug.fr} and \url{https://dx.doi.org/10.17178/EMAA}). EMAA is supported by the Observatoire des Sciences de l’Univers de Grenoble (OSUG).
\end{acknowledgements}

\bibliographystyle{aa}
\bibliography{references.bib}

\begin{appendix}

\section{Electron-impact HeH$^+$ ro-vibrational collisional (de-)excitation data}
\label{sec:appendix_1}

Because HeH$^+$ has a large dipole (1.7~D), dipolar rotational transitions (i.e., those with $\Delta J=\pm 1)$ are expected to have large cross sections dominated by high-partial waves, i.e., long-range interactions. In such a case, the dipolar Coulomb-Born (CB) theory should be a rather good approximation.

Following \cite{Chu74} and \cite{Neuf89}, the CB cross section for a dipole-allowed ro-vibrational excitation $(\upsilon, J) \to (\upsilon', J')$ can be formulated using the Einstein $A$ coefficients as follows {(adopting Gaussian units):
\begin{equation}
    \sigma(\upsilon, J\to \upsilon', J')= \left(\frac{3}{4\pi^2}\right)\frac{\pi}{k^2}A(\upsilon', J' \to \upsilon, J)\left(\frac{3hc^3}{64\pi^4\nu^3}\right)\left(\frac{1}{e^2a_0^2}\right)\frac{(2J'+1)}{(2J+1)}f_{E_1}(\eta, \zeta)
\end{equation}}
where $k$ ($k'$) is the initial (final) wave number of the electron, $A(\upsilon', J' \to \upsilon, J)$ is the Einstein coefficient of the transition, $h$ is the Planck's constant, $c$ is the speed of light, $\nu$ is the frequency of the transition, {$e$ is the elementary charge, $a_0$ is the bohr radius,} and $f_{E1}(\eta, \zeta)$ is a function related to the $E1$ nuclear Coulomb function, with $\eta=-1/k$ and $\zeta=1/k-1/k'$.
The exact expression of $f_{E1}(\eta, \zeta)$ is given in Eq.~(22) of \cite{Chu74}.
We note that the CB approximation leads to a large and finite cross section at the threshold, where the expression of $f_{E1}(\eta, \zeta)$ simplifies \citep{Chu74}. In practice, Einstein coefficients are extracted from the \texttt{EXOMOL} database \citep{exomol16} for dipole-allowed transitions among all {bound levels below a threshold of 14,874~cm$^{-1}$, thus including rotational levels up to $J=23$ and vibrational levels up to $\upsilon=11$. CB cross sections are computed for electron energies up to 20~eV and for 415 dipolar transitions $(\upsilon, J) \to (\upsilon', J')$.} Rate coefficients are deduced for temperatures between 10~K and 20\,000~K by Maxwell-Boltzmann averaging the cross sections.

\begin{figure}
    \centering
    \includegraphics[width=0.49\textwidth]{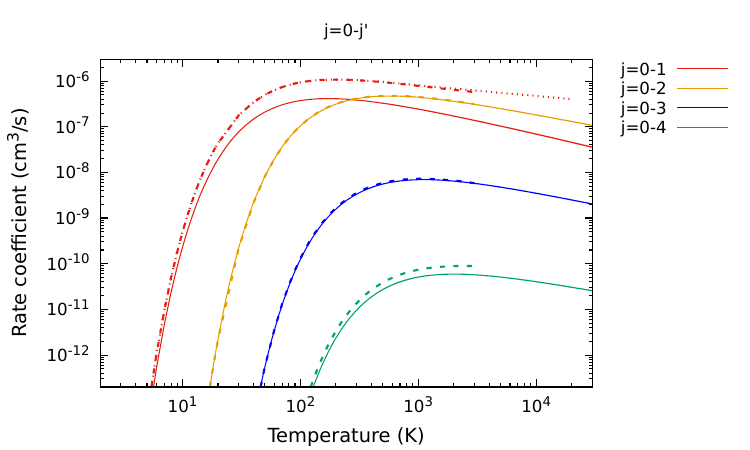}
    \includegraphics[width=0.49\textwidth]{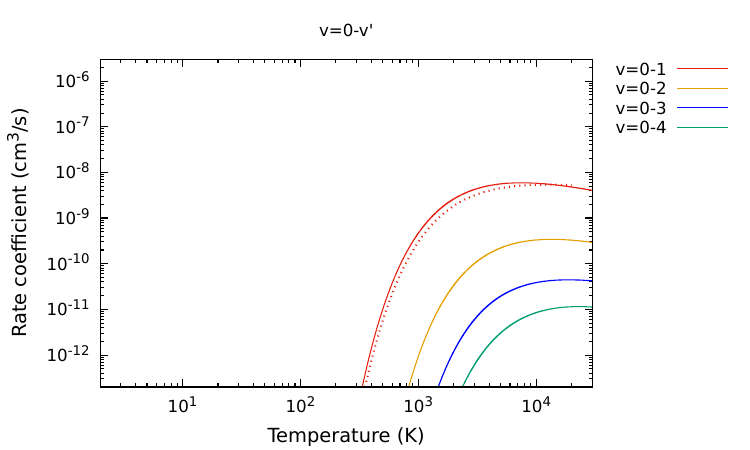}
    \caption{Thermal rate coefficients for several rotational (upper panel) and vibrational (lower panel) excitation transitions of HeH$^+$. Solid lines in both panels are calculated rate coefficients by \cite{ayou19}, dashed lines in the upper panel are the calculations by \cite{hami16}, and dotted lines in both panels indicate our CB data.
    \label{fig:col_rate_heh+}}
\end{figure}

As shown in the upper panel of Fig.~\ref{fig:col_rate_heh+}, the CB rate coefficient for $J=0-1$ is in very good agreement with the result of \cite{hami16}, but it is somewhat larger than that of \cite{ayou19}. This reflects the different treatments of the long-range interaction and the contribution of high partial waves, as discussed in \cite{curik17}. For higher transitions (those with $\Delta J>1$), the dipolar CB approximation predicts null cross sections by definition. The calculations of \cite{hami16} and \cite{ayou19} have shown, however, that the transition $J=0-2$ has a sizable cross-section, which is typically half the cross section for $J=0-1$ at high collision energies. Cross sections for transitions with $\Delta J=\pm3, \pm4$ are, on the other hand, lower by at least an order of magnitude. Similar results are observed for vibrational excitation (lower panel of Fig.~\ref{fig:col_rate_heh+}) where the transition $\upsilon=0-1$ is found to largely dominate with a good agreement between the CB results and those of \cite{ayou19}. As expected, the dipolar CB theory is thus found to be accurate for transitions where $\Delta J=\pm 1$ and $\Delta \upsilon=0, \pm 1$. This confirms that these transitions (in HeH$^+$) are largely dominated by high-partial waves. The next dominant transitions, those with $\Delta J=\pm2$ (and $\Delta \upsilon=0$), are dominated by low-partial waves and cannot be predicted with the CB theory \citep{Faur01}. For these transitions (e.g., $\upsilon=0, J=0\to \upsilon=0, J=2$), the rate coefficients are fixed at half those of the corresponding $\Delta J=\pm1$ transitions (e.g., $\upsilon=0, J=0\to \upsilon=0, J=1$).
This simple recipe shows good results at 500~K when compared to \texttt{RADEX} calculations based on the complete and accurate dataset of \cite{hami16}.
This new extended data file are made available on \href{https://doi.org/10.5281/zenodo.14097180}{Zenodo}.

\section{ArH$^+$ and NeH$^+$ in the dense part of the RS Oph and V1716 Sco ejecta} \label{sec:appendix_2}

\begin{figure*}
    \centering
    \includegraphics[width=0.49\textwidth]{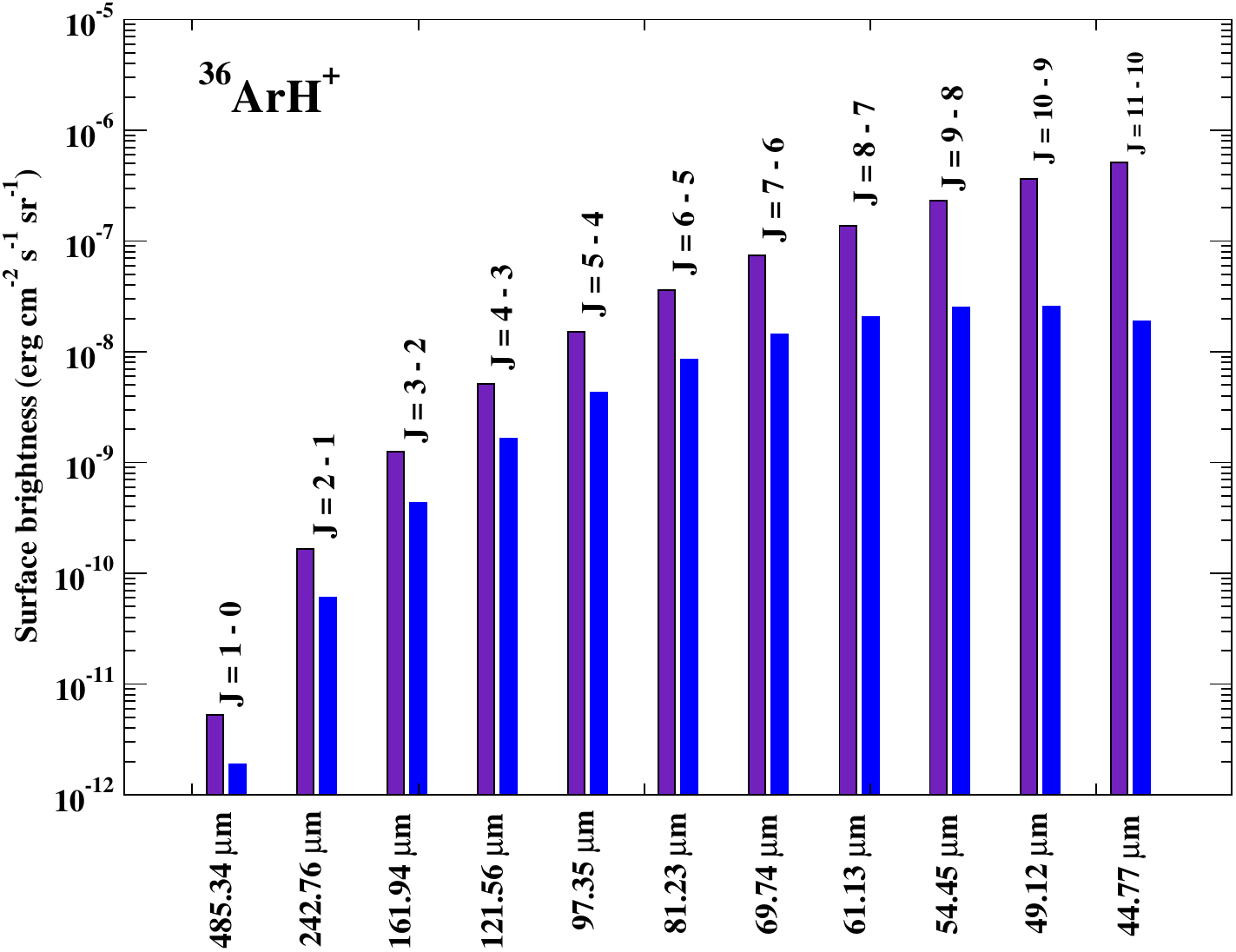}
    \includegraphics[width=0.49\textwidth]{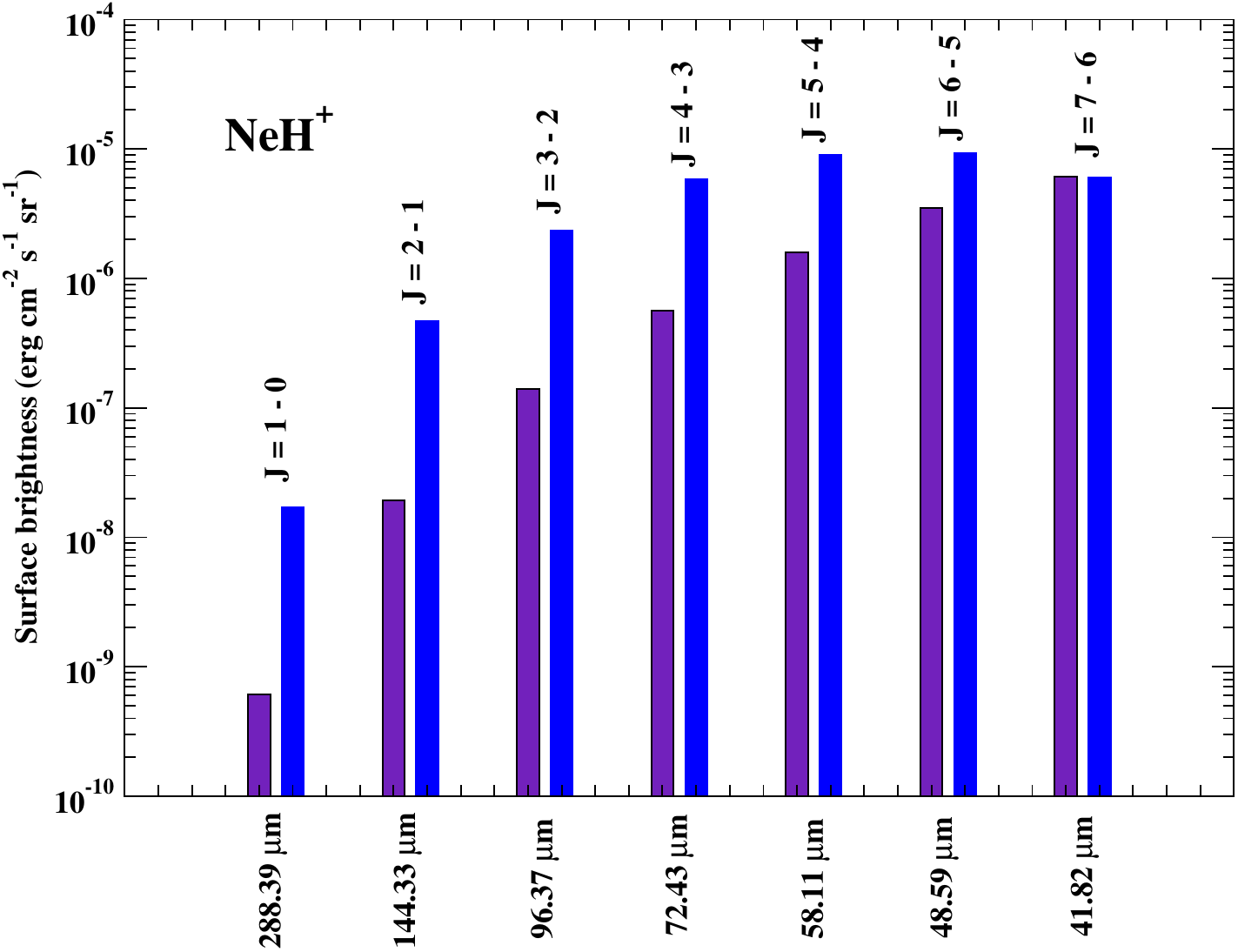}
    \caption{Surface brightness of some potential {$\rm{^{36}ArH^+}$} (left panel) and NeH$^+$ (right panel) rotational transitions in the dense part of the RS Oph (indigo) and V1716 Sco (blue) ejecta.
    The plotted data are summarized in \autoref{tab:transition_RS_Oph}.
    \label{fig:sb_rs_oph}}
\end{figure*}

\autoref{fig:sb_rs_oph} shows the predicted absolute intrinsic line surface brightness of {$\rm{^{36}ArH^+}$} and NeH$^+$ from the dense part of the RS Oph and V1716 Sco ejecta as stated in Sect.~\ref{sec:rs_oph} and \ref{sec:v1716_sco}, respectively.
Surface brightness values along with upper-state energy, frequency, and line-center optical depth of the transitions are noted in \autoref{tab:transition_RS_Oph}.
Since collisional rates for only a limited number of energy levels are available for {$\rm{^{36}ArH^+}$} and NeH$^+$, we do not get any transitions in the JWST domain.
Moreover, since we do not have high-temperature collisional data for {$\rm{^{36}ArH^+}$} and NeH$^+$, these results can be considered as an approximation for the nova outburst region.
The intense surface brightness for {$\rm{^{36}ArH^+}$} and NeH$^+$ transitions arises because of the high density ($\sim10^9$~cm$^{-3}$) of electrons around this region.
From \autoref{fig:sb_rs_oph} and \autoref{tab:transition_RS_Oph}, the most intense transitions of {$\rm{^{36}ArH^+}$} and NeH$^+$ are the $J=11-10$ and $J=7-6$ transitions, respectively for RS Oph and $J=10-9$ and $J=6-5$ transitions, respectively for V1716 Sco.

\begin{table*}
\caption{Strongest transitions of {$\rm{^{36}ArH^+}$} and NeH$^+$ for the dense part in RS Oph and V1716 Sco ejecta.
\label{tab:transition_RS_Oph}}
\centering
\resizebox{\linewidth}{!}{\begin{tabular}{cccccccc}
\hline
{\bf Species} & {\bf Transitions} & {\bf $\rm{E_U/k_B}$} & {\bf Frequency / Wavelength} & \multicolumn{2}{c}{\bf Optical Depth ($\tau$)} & \multicolumn{2}{c}{\bf Surface Brightness (erg~cm$^{-2}$~s$^{-1}$~sr$^{-1}$)} \\
&& {\bf (K)} & {\bf (GHz /~\micron)} & {\bf RS Oph} & {\bf V1716 Sco} & {\bf RS Oph} & {\bf V1716 Sco} \\
\hline
{$\rm{^{36}ArH^+}$} & $\rm{J = 1-0}$ & 29.64 & 617.52 / 485.34 & {$1.59\times10^{-9}$} & {$-8.29\times10^{-10}$} & {$5.29\times10^{-12}$} & {$1.89\times10^{-12}$} \\
{$\rm{^{36}ArH^+}$} & $\rm{J = 2-1}$ & 88.89 & 1234.60 / 242.76 & {$6.34\times10^{-9}$} & {$2.06\times10^{-9}$} & {$1.67\times10^{-10}$} & {$5.99\times10^{-11}$} \\
{$\rm{^{36}ArH^+}$} & $\rm{J = 3-2}$ & 177.71 & 1850.78 / 161.94 & {$1.43\times10^{-8}$} & {$2.55\times10^{-8}$} & {$1.25\times10^{-9}$} & {$4.34\times10^{-10}$} \\
{$\rm{^{36}ArH^+}$} & $\rm{J = 4-3}$ & 296.04 & 2465.62 / 121.56  & {$2.55\times10^{-8}$} & {$7.06\times10^{-8}$} & {$5.14\times10^{-9}$} & {$1.64\times10^{-9}$} \\
{$\rm{^{36}ArH^+}$} & $\rm{J = 5-4}$ & 443.80 & 3078.68 / 97.35 & {$4.05\times10^{-8}$} & {$1.16\times10^{-7}$} & {$1.52\times10^{-8}$} & {$4.24\times10^{-9}$} \\
{$\rm{^{36}ArH^+}$} & $\rm{J = 6-5}$ & 620.86 & 3689.50 / 81.23 & {$5.91\times10^{-8}$} & {$1.45\times10^{-7}$} & {$3.64\times10^{-8}$} & {$8.52\times10^{-9}$} \\
{$\rm{^{36}ArH^+}$} & $\rm{J = 7-6}$ & 827.12 & 4297.65 / 69.74 & {$8.14\times10^{-8}$} & {$1.57\times10^{-7}$} & {$7.49\times10^{-8}$} & {$1.42\times10^{-8}$} \\
{$\rm{^{36}ArH^+}$} & $\rm{J = 8-7}$ & 1062.41 & 4902.68 / 61.13  & {$9.86\times10^{-8}$} & {$1.54\times10^{-7}$} & {$1.38\times10^{-7}$} & {$2.05\times10^{-8}$} \\
{$\rm{^{36}ArH^+}$} & $\rm{J = 9-8}$ & 1326.57 & 5504.16 / 54.45 & {$1.33\times10^{-7}$} & {$1.59\times10^{-7}$} & {$2.33\times10^{-7}$} & {$2.50\times10^{-8}$} \\
{$\rm{^{36}ArH^+}$} & $\rm{J = 10-9}$ & 1619.40 & 6101.65 / 49.12 & {$1.74\times10^{-7}$} & {$1.48\times10^{-7}$} & {$3.62\times10^{-7}$} & {$2.54\times10^{-8}$} \\
{$\rm{^{36}ArH^+}$} & $\rm{J = 11-10}$ & 1940.70 & 6694.72 / 44.77 & {$2.72\times10^{-7}$} & {$1.32\times10^{-7}$} & {$5.10\times10^{-7}$} & {$1.88\times10^{-8}$} \\
\hline
NeH$^+$ & $\rm{J = 1-0}$ & 49.53 & 1039.25 / 288.39 & {$2.87\times10^{-8}$} & {$-6.02\times10^{-6}$} & {$6.14\times10^{-10}$} & {$1.71\times10^{-8}$} \\
NeH$^+$ & $\rm{J = 2-1}$ & 148.50 & 2076.57 / 144.33 & {$1.19\times10^{-7}$} & {$4.85\times10^{-5}$} & {$1.92\times10^{-8}$} & {$4.64\times10^{-7}$} \\
NeH$^+$ & $\rm{J = 3-2}$ & 296.72 & 3110.02 / 96.37 & {$2.87\times10^{-7}$} & {$1.30\times10^{-4}$} & {$1.41\times10^{-7}$} & {$2.35\times10^{-6}$} \\
NeH$^+$ & $\rm{J = 4-3}$ & 493.92 & 4137.67 / 72.43 & {$4.60\times10^{-7}$} & {$1.38\times10^{-4}$} & {$5.66\times10^{-7}$} & {$5.82\times10^{-6}$} \\
NeH$^+$ & $\rm{J = 5-4}$ & 739.73 & 5157.61 / 58.11 & {$9.28\times10^{-7}$} & {$1.17\times10^{-4}$} & {$1.59\times10^{-6}$} & {$8.93\times10^{-6}$} \\
NeH$^+$ & $\rm{J = 6-5}$ & 1033.68 & 6167.92 / 48.59 & {$1.59\times10^{-6}$} & {$8.19\times10^{-5}$} & {$3.51\times10^{-6}$} & {$9.26\times10^{-6}$} \\
NeH$^+$ & $\rm{J = 7-6}$ & 1375.24 & 7166.70 / 41.82  & {$2.96\times10^{-6}$} & {$4.76\times10^{-5}$} & {$6.13\times10^{-6}$} & {$5.99\times10^{-6}$} \\
\hline
\end{tabular}}
\end{table*}

\clearpage

\onecolumn

\section{Reaction network}
\label{sec:appendix_3}

\begin{longtable}{cllc}
\caption{\label{table:reaction_Ar} Reaction pathways for the formation and destruction of noble gas ions considered in \textsc{Cloudy} modeling.} \\
\hline\hline
{\bf Reaction Number} & {\bf Reactions} & {\bf Rate and rate coefficient} & {\bf Comments} \\
{\bf (Type)} &  &  & {\bf and/or references} \\
\hline
\endfirsthead
\caption{continued.}\\
\hline\hline
{\bf Reaction Number} & {\bf Reactions} & {\bf Rate and rate coefficient} & {\bf Comments} \\
{\bf (Type)} &  &  & {\bf and/or references} \\
\hline
\endhead
\hline
\endfoot
\multicolumn{4}{c}{\bf  Ar chemistry} \\
\hline
1 (CR)&$\rm{Ar + CR \rightarrow Ar^{+} + e^{-}}$ & $\rm{10\zeta_{H,cr}} \ s^{-1}$ & {c, p} \\
2 (CRPHOT)&$\rm{Ar + CRPHOT \rightarrow Ar^{+} + e^{-}}$ &$\rm{0.8\frac{\zeta_{H_2,cr}}{1-\omega}} \ s^{-1}$ & {c, p} \\
3 (IN)&$\rm{Ar + H_2^{+} \rightarrow ArH^{+} + H}$ &$\rm{10^{-9}}\ cm^3\ s^{-1} $ & {s, n1} \\
4 (IN)&$\rm{Ar + H_3^{+} \rightarrow ArH^{+} + H_2}$&$\rm{8\times10^{-10}exp\left(\frac{-6019~K}{T}\right)} \ cm^3\ s^{-1}$ & {v, d} \\
5 (IN)&$\rm{Ar^{+} + H_2 \rightarrow ArH^{+} + H}$&$\rm{8.4\times10^{-10}\left(\frac{T}{300~K}\right)^{0.16}} \ cm^3\ s^{-1}$ & {s} \\
6 (IN)&$\rm{ArH^{+} + H_2 \rightarrow Ar + H_3^{+}}$&$\rm{8\times10^{-10}}\ cm^3\ s^{-1}$ & {v} \\
7 (IN)&$\rm{ArH^{+} + CO \rightarrow Ar + HCO^{+}}$&$\rm{1.25\times10^{-9}}\ cm^3\ s^{-1}$ & {v} \\
8 (IN)&$\rm{ArH^{+} + O \rightarrow Ar + OH^{+}}$&$\rm{8\times10^{-10}}\ cm^3\ s^{-1}$ & {s} \\
9 (IN)&$\rm{ArH^{+} + C \rightarrow Ar + CH^{+}}$&$\rm{8\times10^{-10}}\ cm^3\ s^{-1}$ & {s} \\
10 ({CT})&$\rm{Ar^{++} + H \rightarrow Ar^{+} + H^{+}}$&$\rm{10^{-15}}\ cm^3\ s^{-1}$ & {k1} \\
11 ({CT})&$\rm{Ar + N_2^+ \rightarrow Ar^+ + N_2}$&$\rm{3.65\times10^{-10}}\ cm^3\ s^{-1}$ & {c}  \\
12 ({CT})&$\rm{Ar^+ + H_2 \rightarrow Ar + H_2^+}$&$\rm{2.00\times10^{-12}}\ cm^3\ s^{-1}$ & {c}  \\
13 ({CT})&$\rm{Ar^+ + O_2 \rightarrow Ar + O_2^+}$&$\rm{3.50\times10^{-11}}\ cm^3\ s^{-1}$ & {c} \\
14 (IN)&$\rm{Ar^+ + CH_4 \rightarrow CH_2^+ + Ar + H_2}$&$\rm{1.40\times10^{-10}}\ cm^3\ s^{-1}$ & {c}  \\
15 (IN)&$\rm{Ar^+ + CH_4 \rightarrow CH_3^+ + Ar + H}$&$\rm{7.90\times10^{-10}}\ cm^3\ s^{-1}$ & {c} \\
16 ({CT})&$\rm{Ar^+ + HCl \rightarrow Ar + HCl^+}$&$\rm{2.90\times10^{-10}}\ cm^3\ s^{-1}$ & {c} \\
17 (IN)&$\rm{Ar^+ + HCl \rightarrow ArH^+ + Cl}$&$\rm{6.00\times10^{-11}}\ cm^3\ s^{-1}$ & {c} \\
18 ({CT})&$\rm{Ar^+ + CO \rightarrow Ar + CO^+}$&$\rm{2.80\times10^{-11}}\ cm^3\ s^{-1}$ & {c} \\
19 ({CT})&$\rm{Ar^+ + NH_3 \rightarrow Ar + NH_3^+}$&$\rm{1.60\times10^{-9}}\ cm^3\ s^{-1}$ & {c} \\
20 ({CT})&$\rm{Ar^+ + N_2 \rightarrow Ar + N_2^+}$&$\rm{1.20\times10^{-11}}\ cm^3\ s^{-1}$ & {c} \\
21 ({CT})&$\rm{Ar^+ + H_2O \rightarrow Ar + H_2O^+}$&$\rm{1.30\times10^{-9}}\ cm^3\ s^{-1}$ & {c} \\
22 (XR)&$\rm{Ar + XR \rightarrow Ar^{++} + e^{-} + e^{-}}$ & {---} & {c} \\
23 (XR)&$\rm{Ar^{+} + XR \rightarrow Ar^{++} + e^{-}}$& {---} & {c} \\
24 (XRSEC)&$\rm{Ar + XRSEC \rightarrow Ar^{+} + e^{-}}$& {---} & {c}
 \\
25 (XRPHOT)&$\rm{Ar + XRPHOT \rightarrow Ar^{+} + e^{-}}$& {---} & {c} \\
26 (ER)&$\rm{Ar^+ + e^{-} \rightarrow Ar + h\nu}$& ---   & {c} \\
27 (ER)&$\rm{Ar^{++} + e^{-} \rightarrow Ar^{+} + h\nu}$& ---  & {c} \\
28 (DR)&$\rm{ArH^{+} + e^{-} \rightarrow Ar + H}$& {$\rm{5.06\times10^{-13} \left(\frac{T}{300~K}\right)^{2.084} exp\left(\frac{-7540~K}{T}\right)}\ cm^3\ s^{-1}$} & {k2} \\
29 (PH)&$\rm{ArH^{+} + h\nu \rightarrow Ar^+ + H}$&$\rm{4.20\times10^{-12}exp(-3.27A_v)\ s^{-1}}$ & {r} \\
\hline
\multicolumn{4}{c}{\bf  Ne chemistry}\\
\hline
1 (CR)&$\rm{Ne + CR \rightarrow Ne^{+} + e^{-}}$ & $\rm{10\zeta_{H,cr}}\ s^{-1}$ & {c, d} \\
2 (CRPHOT)&$\rm{Ne + CRPHOT \rightarrow Ne^{+} + e^{-}}$ &$\rm{0.8\frac{\zeta_{H_2,cr}}{1-\omega}}\ s^{-1}$ & {c, d} \\
3 (IN)&$\rm{Ne + H_2^{+} \rightarrow NeH^{+} + H}$ & $\rm{2.58\times10^{-10}exp\left(\frac{-6717~K}{T}\right)}\ cm^3\ s^{-1}$ & {d} \\
4 (IN)&$\rm{Ne + H_3^{+} \rightarrow NeH^{+} + H_2}$&$\rm{8\times10^{-10}exp\left(\frac{-27456~K}{T}\right)}\ cm^3\ s^{-1}$ & {d} \\
5 (IN) & $\rm{Ne^{+} + H_2 \rightarrow Ne + H + H^+}$ & $\rm{1.98\times10^{-14}exp\left(\frac{-35~K}{T}\right)}$ cm$^3$ s$^{-1}$  & {d} \\
6 ({CT}) & $\rm{Ne^{+} + H_2 \rightarrow Ne + H_2^+}$ & $\rm{4.84\times10^{-15}}$ cm$^3$ s$^{-1}$ & {d}  \\
7 (IN)&$\rm{NeH^{+} + H_2 \rightarrow Ne + H_3^{+}}$&$\rm{3.65\times10^{-9}}\ cm^3\ s^{-1}$ & {d} \\
8 (IN)&$\rm{NeH^{+} + CO \rightarrow Ne + HCO^{+}}$&$\rm{2.26\times10^{-9}}\ cm^3\ s^{-1}$ & {d} \\
9 (IN)&$\rm{NeH^{+} + O \rightarrow Ne + OH^{+}}$&$\rm{2.54\times10^{-9}}\ cm^3\ s^{-1}$ & {d} \\
10 (IN)&$\rm{NeH^{+} + C \rightarrow Ne + CH^{+}}$&$\rm{1.15\times10^{-9}}\ cm^3\ s^{-1}$ & {d} \\
11 ({CT})&$\rm{Ne^{++} + H \rightarrow Ne^{+} + H^{+}}$&$\rm{1.94\times10^{-15}}\ cm^3\ s^{-1}$ & {d} \\
12 (IN)&$\rm{HeH^+ + Ne \rightarrow NeH^+ +He }$&$\rm{1.25 \times 10^{-9}}\ cm^3\ s^{-1}$ & {c} \\
13 (IN)&$\rm{NeH^+ + He\rightarrow HeH^{+} + Ne }$&$\rm{3.8 \times 10^{-14}}\ cm^3\ s^{-1}$  & {c} \\
14 (IN)&$\rm{Ne^{+} + CH_4 \rightarrow CH^{+} + Ne + H_2+ H}$&$\rm{8.4\times10^{-13}}\ cm^3\ s^{-1}$ & {c} \\
15 (IN)&$\rm{Ne^{+} + CH_4 \rightarrow {CH_2}{^+} + Ne + H_2}$&$\rm{4.2\times10^{-12}}\ cm^3\ s^{-1}$ & {c}  \\
16 (IN)&$\rm{Ne^{+} + CH_4 \rightarrow {CH_3}{^+} + Ne + H}$&$\rm{4.7\times10^{-12}}\ cm^3\ s^{-1}$ & {c} \\
17 ({CT})&$\rm{Ne^{+} + CH_4 \rightarrow {CH_4}{^+} + Ne}$&$\rm{1.1\times10^{-11}}\ cm^3\ s^{-1}$ & {c} \\
18 (IN)&$\rm{Ne^{+} + NH_3 \rightarrow {NH}{^+} + Ne+H_2}$&$\rm{4.5\times10^{-12}}\ cm^3\ s^{-1}$ & {c} \\
19 (IN)&$\rm{Ne^{+} + NH_3 \rightarrow {NH_2}{^+} + Ne+H}$&$\rm{1.9\times10^{-10}}\ cm^3\ s^{-1}$ & {c} \\
20 ({CT})&$\rm{Ne^{+} + NH_3 \rightarrow {NH_3}{^+} + Ne}$&$\rm{2.7\times10^{-11}}\ cm^3\ s^{-1}$ & {c} \\
21 ({CT})&$\rm{Ne^{+} + N_2 \rightarrow {N_2}{^+} + Ne}$&$\rm{1.1\times10^{-13}}\ cm^3\ s^{-1}$ & {c} \\
22 ({CT})&$\rm{Ne^{+} + H_2O \rightarrow {H_2O}{^+} + Ne}$&$\rm{8.0\times10^{-10}}\ cm^3\ s^{-1}$ & {c} \\
23 (IN)&$\rm{Ne^{+} + O_2 \rightarrow {O}{^+} + Ne + O}$&$\rm{6.0\times10^{-11}}\ cm^3\ s^{-1}$ & {c} \\
24 (XR)&$\rm{Ne + XR \rightarrow Ne^{++} + e^{-} + e^{-}}$&{---} & {c} \\
25 (XR)&$\rm{Ne^{+} + XR \rightarrow Ne^{++} + e^{-}}$&{---} & {c} \\
26 (XRSEC)&$\rm{Ne + XRSEC \rightarrow Ne^{+} + e^{-}}$&{---} & {c} \\
27 (XRPHOT)&$\rm{Ne + XRPHOT \rightarrow Ne^{+} + e^{-}}$&{---} & {c} \\
28 (ER)&$\rm{Ne^+ + e^{-} \rightarrow Ne + h\nu}$& ---  & {c} \\
29 (ER)&$\rm{Ne^{++} + e^{-} \rightarrow Ne^{+} + h\nu}$&  --- & {c} \\
30 (DR)&$\rm{NeH^{+} + e^{-} \rightarrow Ne + H}$& {$\rm{5.06\times10^{-13} \left(\frac{T}{300~K}\right)^{2.084} exp\left(\frac{-7540~K}{T}\right)}\ cm^3\ s^{-1}$} & {`guessed' following k2} \\
31 (PH)&$\rm{NeH^{+} + h\nu \rightarrow Ne^{+} + H}$&$\rm{4.20\times10^{-12}exp(-3.27A_v)\ s^{-1}}$ &  {d} \\
\hline
\multicolumn{3}{c}{\bf  He chemistry}\\
\hline
1 (CR)&$\rm{He + CR \rightarrow He^{+} + e^{-}}$ & $\rm{10\zeta_{H,cr}}\ s^{-1}$ & {c, d} \\
2 (CRPHOT)&$\rm{He + CRPHOT \rightarrow He^{+} + e^{-}}$ &$\rm{0.8\frac{\zeta_{H_2,cr}}{1-\omega}} \ s^{-1}$ & {c, d} \\
3 (IN)&$\rm{He + H_2^{+} \rightarrow HeH^{+} + H}$ & $\rm{3\times10^{-10}exp\left(\frac{-6717\ K}{T}\right)}\ cm^3\ s^{-1}$ &  {b} \\
4 (IN)&$\rm{He + H_3^{+} \rightarrow HeH^{+} + H_2}$&$\rm{8\times10^{-10}exp\left(\frac{-29110\ K}{T}\right)}\ cm^3\ s^{-1}$ &  {d} \\
5 (IN) & $\rm{He^{+} + H_2 \rightarrow He + H + H^+}$ & $\rm{3.70\times10^{-14}exp\left(\frac{-35~K}{T}\right)}$ cm$^3$ s$^{-1}$ & {c} \\
6 ({CT}) & $\rm{He^{+} + H_2 \rightarrow He + H_2^+}$ & $\rm{7.20\times10^{-15}}$ cm$^3$ s$^{-1}$ & {c} \\
7 (IN)&$\rm{HeH^{+} + H_2 \rightarrow He + H_3^{+}}$&$\rm{1.26\times10^{-9}}\ cm^3\ s^{-1}$ & {o} \\
8 (IN)&$\rm{HeH^{+} + CO \rightarrow He + HCO^{+}}$&$\rm{2.33\times10^{-9}}\ cm^3\ s^{-1}$ & {d} \\
9 (IN)&$\rm{HeH^{+} + O \rightarrow He + OH^{+}}$&$\rm{2.68\times10^{-9}}\ cm^3\ s^{-1}$ & {d} \\
10 (IN)&$\rm{HeH^{+} + C \rightarrow He + CH^{+}}$&$\rm{1.18\times10^{-9}}\ cm^3\ s^{-1}$ & {d} \\
11 ({CT})&$\rm{He^{++} + H \rightarrow He^{+} + H^{+}}$&$\rm{2.45\times10^{-15}}\ cm^3\ s^{-1}$ & {d} \\
12 (IN) &$\rm{HeH^{+} + H \rightarrow He + H_2^{+}}$&$\rm{1.7\times10^{-9}}\ cm^3\ s^{-1}$  & {e} \\
13 (RA) &$\rm{He^+ + H \rightarrow HeH^{+} + h\nu}$ &$\rm{1.44\times10^{-16}}\ cm^3\ s^{-1}$ & {g, n3} \\
14 (RA) &$\rm{He + H^+ \rightarrow HeH^{+} + h\nu}$ & $\rm{5.6\times10^{-21}\left(\frac{T}{10^4K}\right)^{-1.25}}\ cm^3\ s^{-1}$ & {n3} \\
15 (XR)&$\rm{He + XR \rightarrow He^{++} + e^{-} + e^{-}}$ & {---} & {c} \\
16 (XR)&$\rm{He^{+} + XR \rightarrow He^{++} + e^{-}}$&{---} & {c} \\
17 (XRSEC)&$\rm{He + XRSEC \rightarrow He^{+} + e^{-}}$&{---} & {c} \\
18 (XRPHOT)&$\rm{He + XRPHOT \rightarrow He^{+} + e^{-}}$&{---} & {c} \\
19 (ER)&$\rm{He^+ + e^{-} \rightarrow He + h\nu}$&  --- & {c} \\
20 (ER)&$\rm{He^{++} + e^{-} \rightarrow He^{+} + h\nu}$&  --- & {c} \\
21 (DR)&$\rm{HeH^{+} + e^{-} \rightarrow He + H}$& $\rm{4.3\times10^{-10}\left(\frac{T}{10^4K}\right)^{-0.5}}\ cm^3\ s^{-1}$ & {n2} \\
22 (PH)&$\rm{HeH^{+} + h\nu \rightarrow He^{+} + H}$& --- & {c, n3}  \\
23 ({AI}) & $\rm{He^+ + H^- \rightarrow HeH^+ + e^-}$ & $\rm{3.2\times10^{-11}\left(\frac{T}{10^4K}\right)^{-0.34}}\ cm^3\ s^{-1}$ & {n3} \\
\hline
\multicolumn{4}{c}{\bf  Additional modified chemistry}\\
\hline
1 (RA) & $\rm{H^+ + H \rightarrow H_2^+ + h\nu}$ & $\rm{2.3\times10^{-16}\left(\frac{T}{10^4K}\right)^{1.5}}\ cm^3\ s^{-1}$ & {c, n3} \\
2 (DR) & $\rm{H_2^+ + e^- \rightarrow H + H}$ & $\rm{3\times10^{-9}\left(\frac{T}{10^4K}\right)^{-0.4}}\ cm^3\ s^{-1}$ & {c, n3} \\
\end{longtable}
\tablefoot{CR refers to cosmic rays, CRPHOT to secondary photons produced by cosmic rays, XR to direct X-rays, XRSEC to secondary electrons produced by X-rays, XRPHOT to secondary photons from X-rays, IN to ion-neutral reactions, {CT to charge transfer reactions, AI to associative ionization reactions,} RA to radiative association reactions, ER to electronic recombination reactions for atomic ions, DR to dissociative recombination reactions for molecular ions, PH to photodissociation reactions, h$\nu$ to a photon, $\zeta$ to cosmic-ray or X-ray ionization
rates, and $\omega$ is the dust albedo. \\
{\textsuperscript{c} Reaction pathways are already included and/or reaction rates are self-consistently controlled in \textsc{Cloudy} by default. \\
\textsuperscript{p} \citet[and references therein]{prie17}, \textsuperscript{s} \citet[and references therein]{schi14}, \textsuperscript{n1} \cite{neuf16}, \textsuperscript{v} \cite{vill82}, \textsuperscript{d} \cite{das20}, \textsuperscript{k1} \cite{king96}, \textsuperscript{k2} \citet[valid in the temperature intervals of $7000-20000$~K]{kalo24}, \textsuperscript{r} \cite{roue14}, \textsuperscript{o} \cite{orie77}, \textsuperscript{e} \cite{espo15}, \textsuperscript{n2} \cite{novo19}, \textsuperscript{b} \cite{blac78}, \textsuperscript{g} \citet[and references therein]{gust19}, \textsuperscript{n3} \citet[and references therein]{neuf20}
}
}

\end{appendix}

\end{document}